\documentclass{aa}  

\usepackage{graphicx}
\usepackage{txfonts}
\usepackage{lipsum}
\usepackage{subcaption}         % necessary for continued figures, example in section 3
\usepackage{lscape}             % to rotate a single page table, example in appendix.
\usepackage{placeins}           % useful with \FloatBarrier, to keep 
\usepackage{booktabs}
\usepackage{threeparttable}
\newif\ifshowchanges
\showchangesfalse
\newcommand{\change}[1]{%
  \ifshowchanges
    \textbf{#1}%
  \else
    #1%
  \fi
}

\newcommand{\diff}{\ensuremath{\mathrm{d}}}% creates the correct unit font in math 

\begin{document}
\def\aj{Astron. J.}
\def\actaa{Acta Astron.}
\def\araa{Annu. Rev. Astron. Astrophys.}
\def\apj{Astrophys. J.}
\def\apjl{Astrophys. J. Lett.}
\def\apjs{Astrophys. J. Suppl.}
\def\ao{Appl. Opt.}
\def\apss{Astrophys. Space Sci.}
\def\aap{Astron. Astrophys.}
\def\aapr{Astron. Astrophys. Rev.}
\def\aaps{Astron. Astrophys. Suppl.}
\def\azh{Astronomicheskii Zhurnal}
\def\baas{Bulletin of the AAS}
\def\bac{Bulletin of the Astronomical Institutes of Czechoslovakia}
\def\caa{Chinese Astronomy and Astrophysics}
\def\cjaa{Chinese Journal of Astronomy and Astrophysics}
\def\icarus{Icarus}
\def\jcap{JCAP}
\def\jrasc{Journal of the RAS of Canada}
\def\memras{Memoirs of the RAS}
\def\mnras{Mon. Not. Roy. Astron. Soc.}
\def\na{New Astronomy}
\def\nar{New Astronomy Review}
\def\pra{Phys. Rev. A: General Physics}
\def\prb{Phys. Rev. B: Solid State}
\def\prc{Phys. Rev. C}
\def\prd{Phys. Rev. D}
\def\pre{Phys. Rev. E}
\def\prl{Phys. Rev. Lett.}
\def\pasa{Publications of the Astron. Soc. of Australia}
\def\pasp{Publications of the ASP}
\def\pasj{Publications of the ASJ}
\def\rmxaa{Revista Mexicana de Astronomia y Astrofisica}
\def\qjras{Quarterly Journal of the RAS}
\def\skytel{Sky and Telescope}
\def\solphys{Solar Physics}
\def\sovast{Soviet Astronomy}
\def\ssr{Space Science Reviews}
\def\zap{Zeitschrift für Astrophysik}
\def\nat{Nature}
\def\iaucirc{IAU Cirulars}
\def\aplett{Astrophys. Lett.}
\def\apspr{Astrophysics Space Physics Research}
\def\bain{Bulletin Astronomical Institute of the Netherlands}
\def\fcp{Fundamental Cosmic Physics}
\def\gca{Geochimica Cosmochimica Acta}
\def\grl{Geophysics Research Letters}
\def\jcp{Journal of Chemical Physics}
\def\jgr{Journal of Geophysics Research}
\def\jqsrt{Journal of Quantitiative Spectroscopy and Radiative Transfer}
\def\memsai{Mem. Societa Astronomica Italiana}
\def\nphysa{Nuclear Physics A}
\def\physrep{Physics Reports}
\def\physscr{Physica Scripta}
\def\planss{Planetary Space Science}
\def\procspie{Proceedings of the SPIE}

\let\astap=\aap
\let\apjlett=\apjl
\let\apjsupp=\apjs
\let\applopt=\ao

\def\ndash{--}
\def\mdash{---}

   \title{Bayesian parameter study of the Seyfert-starburst composite galaxies NGC\,1068 and NGC\,7469}

   %\subtitle{Subtitle}

%%%%%%%%%%%%%%%%%%%%%%%%%%%%%%%%%%%%%%%%
% Please do not include ORCIDs next to author names.
% Only ORCIDs authenticated by individual authors in EDP Sciences editorial system will be taken into account.
% ORCIDs included here will be removed.
%%%%%%%%%%%%%%%%%%%%%%%%%%%%%%%%%%%%%%%%
   \author{
   B.~Eichmann \inst{1, 2}
   \and
   S.~Salvatore \inst{1, 2}
   \and 
   S. del Palacio \inst{3}
   \and
   G.~Sommani \inst{2, 4}
   \and 
   C. Mele \inst{2, 4}
   \and
   P. M. Veres \inst{2, 4}
   \and
   J.~Becker Tjus \inst{1, 2, 3}
    }

   \institute{Ruhr-Universit\"at Bochum, Fakult\"at f\"ur Physik und Astronomie, Theoretische Physik IV, 44780 Bochum, Germany   \and Ruhr Astroparticle and Plasma Physics Center (RAPP Center), 44780 Bochum, Germany  \and Department of Space Earth and Environment, Chalmers University of Technology, 412 96 Gothenburg, Sweden \and Ruhr-Universit\"at Bochum, Fakult\"at f\"ur Physik und Astronomie, Astronomisches Institut (AIRUB), 44780 Bochum, Germany}

   \date{Received February 17, 2026}

% \abstract{}{}{}{}{}
% 5 {} token are mandatory
 
  \abstract
   {Multimessenger observation of the Seyfert-starburst composite galaxies NGC\,1068 and NGC\,7469 indicate a characteristic feature in the radio band (the so-called mm-bump) as well as indication of high-energy neutrinos by the AGN corona. Moreover, also the starburst ring of these sources is bright in the radio and hence, a potential source of $\gamma$-rays and neutrinos.}
   {We aim to explain the non-thermal features of these two sources with our homogeneous steady-state Seyfert-starburst composite model, which we refined in this work. Hereby, we account for stochastic diffuse acceleration and energy losses within the corona and $\gamma\gamma$-pair attenuation of the escaping $\gamma$-rays.} 
   {Since the non-thermal features of these Seyfert sources make a minor contribution in the electromagnetic spectrum, we are left with just a few data points that can be attributed to the starburst ring or the AGN corona. Hence, a proper inclusion of the prior information on the physical parameters is needed and subsequently used in the context of a Markov Chain Monte Carlo approach to avoid overfitting.}
   {Based on this Bayesian parameter study we show, that the non-thermal features of NGC\,1068 can be explained well. Still a more detailed treatment of the spatial inhomogeneities in the central region of the AGN could further improve the fit results. This manifests itself even more clearly in the case of NGC\,7469, where the mm-bump needs to emerge from a coronal size $R_{\rm c}>100\,\mathcal{R}_{\rm s}$ \change{(with the Schwarzschild radius $\mathcal{R}_{\rm s}$)}, whereas (TeV-PeV)-neutrino emission requires $R_{\rm c}< 10\,\mathcal{R}_{\rm s}$.}
   {Similar to what has previously been shown in other wavebands, our analysis highlights that the spatial extension of the so-called AGN corona depends the considered energy of the messenger. Hence, it seems that there is not a unique edge of the corona and a substantial progress in the understanding of these phenomena is expected if future analysis account for these spatial inhomogeneities.}


   \maketitle
%%%To switch off the line numbers:
\nolinenumbers

%%%%%%%%%%%%%%%%%%%%%%%%%%%%%%%%%%%%%%%%%%%%%%%%%%%%%%%%%%%%%%
\section{Introduction}
Multimessenger astrophysics has significantly enlarged our understanding of the Universe. Starting with the detection of high-energy charged particles and photons across various wavelengths, an important messenger has been added about a decade ago with the observation of high-energy cosmic neutrinos \citep{IceCube2013_PhRvL}. 

High-energy neutrino production is typically associated with gamma-ray emission; therefore, the known diffuse gamma-ray background sets an upper limit on potential gamma-ray transparent neutrino sources. Based on this constraint, it has been shown \citep[e.g.][]{Murase+2016} that a significant population of so-called ``hidden'' neutrino sources, which are opaque to GeV--TeV gamma rays, is necessary. A promising scenario to achieve this gamma-ray opaqueness is the presence of a bright UV/X-ray photon field close to the neutrino source to absorb GeV--TeV gamma rays through electron-positron pair production. For an active galactic nucleus (AGN) such a photon target is typically present very close (about some tens of Schwarzschild radii) to its central supermassive black hole in the so-called AGN corona, where X-ray photons result from Comptonization of photons emitted by the accretion disk.

Thus, high-energy neutrino sources are not necessarily $\gamma$-ray bright. But nevertheless, the IceCube Collaboration found, with a global significance of $4.2\sigma$, a high-energy neutrino signal at TeV energies from the direction of NGC\,1068 \citep{IceCube2022_Sci} based on a dedicated catalog of 110 astronomical objects, that are bright in the \textit{Fermi}-LAT energy band. Consequently, NGC\,1068 is bright at GeV gamma-rays, but TeV gamma-rays are absent \citep{Acciari+2019}. 
NGC\,1068 is one of the closest Seyfert 2 galaxies, which is a radio-quiet type of an AGN, and further, it hosts a circumnuclear starburst ring. 
Hence, the actual origin of the observed GeV gamma-rays is unclear and a proper understanding of the high-energy signal needs to account for the broadband emission of the different kinds of messengers. To date, different groups \citep[e.g.][]{Inoue+2020, Murase+2020, Kheirandish:2021wkm, Fiorillo+2024} have proposed a model for the AGN corona to address those high-energy neutrinos. The first theoretical model that included all non-thermal emission features of NGC\,1068 has been provided by  \cite{eichmann2022solving}, hereafter referred to as E+22 model. Therein, a spatially homogeneous, steady-state AGN corona and a circumnuclear starburst region are considered that are independent of each other and fueled with cosmic rays (CRs) from individual acceleration sites. It has been shown that the observational features can be nicely explained if both emission sites are taken into account. A rather large CR pressure of $\gtrsim10\%$ of the thermal gas pressure is needed in the AGN corona, such as recently found by \cite{mutie2025consistent}. Moreover, the resulting spectral behavior of the non-thermal protons has been much softer than expected from stochastic diffusive acceleration. 

Most recently \cite{sommani2025two} found two IceCube real-time track alerts, out of just $\sim30$ per year from the direction of NGC\,7469, and estimated a chance coincidence probability of 0.08\%~(referring to a $3.2\,\sigma$ significance of a detection).
The spectral dependence of the neutrino flux is unclear due to the limited number of neutrinos and their energy uncertainty. Still, it seems that those neutrinos are emitted at significantly higher energies than in the case of NGC\,1068. Moreover, NGC\,7469 does not show any significant gamma-ray counterpart---not even at GeV or sub-GeV energies. But such as NGC\,1068, also NGC\,7469 is a Seyfert-starburst composite galaxy, where the circumnuclear starburst ring is responsible for most of the radio emission. However, its central AGN is classified as Seyfert 1 (or sometimes also 1.5) indicating no (or less) attenuation of the central electromagnetic emission by the dusty torus. Thus, also in case of NGC\,7469 the AGN corona constitutes the most likely origin of the potential high-energy neutrino signal. But can those be explained under consideration of the constraints given from its broadband electromagnetic emission?  

In this work, we update the E+22 model and use a Bayesian approach to analyze the parameter space to explain the non-thermal multimessenger emission of NGC\,1068 and NGC\,7469. The paper is structured as follows: in Sect.~\ref{sec:Seyfert-starburst-model}, we introduce the updated Seyfert-starburst composite model, and in Sect.~\ref{sec:methods} we explain the methods used for the comparison with the observational data of NGC\,1068 and NGC\,7469. In Sect.~\ref{sec:results} we show the results, before we conclude and discuss them in Sect.~\ref{sec:conclusions}.

%%%%%%%%%%%%%%%%%%%%%%%%%%%%%%%%%%%%%%%%%%%%%%%%%%%%%%%%%%%%%%%%%%%%%%%%%%%%%%%%%%%%%%%
%%%%%%%%%%%%%%%%%%%%%%%%%%%%%%%%%%%%%%%%%%%%%%%%%%%%%%%%%%%%%%%%%%%%%%%%%%%%%%%%%%%%%%%
%%%%%%%%%%%%%%%%%%%%%%%%%%%%%%%%%%%%%%%%%%%%%%%%%%%%%%%%%%%%%%%%%%%%%%%%%%%%%%%%%%%%%%%
\section{Seyfert-starburst composite model} \label{sec:Seyfert-starburst-model}
Seyfert-starburst composite galaxies are characterized by an AGN core and a surrounding starburst environment which both can significantly contribute to the observed non-thermal emission features. Therefore, an holistic explanation of their phenomena needs to account for both environments. Due to the large spatial scale separation the AGN corona (on about milliparsec scales or less) and the kiloparsec scale starburst ring can be considered as a first approximation as independent of each other. Moreover, we assume a constant strength of the magnetic field (that is uniform on small scales and randomly orientated on significantly larger ones), as well as a thermal gas with a constant density $n_{\rm gas}$ and temperature $T_{\mathsf{gas}}$ in the two environments. 
With respect to the non-thermal emission signatures the most relevant thermal photon targets are the UV photons from the accretion disk and the X-rays of the AGN corona, as well as the infrared (IR) emission from the re-scattered starlight by dust grains in the starburst ring (see Sect.~\ref{sec:methods} for more details). In both cases, it is adopted that these thermal fields are isotropic and stationary, even though the coronal X-ray emission often shows variability on hour and sub-hour scales. 
Ignoring this variability yields some inaccuracy on the coronal photon target that however, will be further discussed in Sect.~\ref{sec:conclusions}. 
%Using an averaged coronal photon density in the following already bares a source of uncertainty that however, will be further discussed in Sect.~\ref{sec:conclusions}. 

Further, also the non-thermal particle population (and its resulting non-thermal emission features) is for mathematical convenience adopted to be spatially homogeneous and in a steady state.\\
These basic assumption have already been used in the E+22 model, but in addition we also account for the following: 
\begin{enumerate}
    \item[(A1)] In the AGN corona, the particle acceleration by stochastic diffusive acceleration (SDA) occurs in the same environment as the energy losses that produce the non-thermal emission features.
    \item[(A2)] In addition to a first generation of $\mathrm{e}^\pm$ from hadronic processes, we include secondaries generated from leptonic radiation processes --- mainly synchrotron and inverse Compton (IC).
    \item[(A3)] For the AGN corona, $\gamma\gamma$-pair attenuation can also occur outside of the corona (by the anisotropic, external X-ray photons that radially leave the corona) and in addition, an increased $\gamma$-ray opaqueness is possible for a shrunken $\gamma$-ray production site that does not align with the X-ray source.
\end{enumerate}
%Hence, the starburst ring is still treated as in E+22 using an accelerator whose escaping CR rate is used as the primary source within this environment. 
To determine the momentum distribution of the relativistic electrons ($i=\mathrm{e}$), positrons ($i=\mathrm{e}^+$) and protons ($i=\mathrm{p}$), we use the diffusion-advection equation in momentum ($p$) space and assume an assigned magnetic turbulence with a power spectrum $W(k)\propto k^{-\varkappa}$ and a strength $\eta^{-1}\equiv 8\pi \int W(k)\,\text{d}k/B^2$. Further, we incorporate the impact of continuous momentum losses ($\langle\dot{p}\rangle<0$) by interactions with, e.g., ambient gas and photon targets, as well as catastrophic losses on a momentum-dependent timescale $t_{\mathrm{esc}}(p)$ such as by escape
from the physical system.\footnote{All details on the energy losses, the escape and the acceleration timescales can be found in E+22.} When the turbulent magnetic field is small compared to the regular magnetic field ($\eta^{-1} < 1$), the steady-state transport equation for the differential CR particle density $n_{i}(p)=4\pi\,p^2\,\langle f(p) \rangle$ in this region becomes
\begin{align}
   -&\frac{\partial}{\partial p}\left[p^2\, D(p) \frac{\partial}{\partial p}\left( \frac{n_{i}}{p^2}\right)\right] \label{eq:momDiffTerm}\\ 
   &+\frac{\partial}{\partial p}\left(\langle\dot{p}\rangle n_{i}\right)+\frac{n_{i}}{t_{\mathrm{esc}}(p)} %\nonumber\\ 
   =
    \begin{cases}
    q_{\mathrm{p}}(p) \, &\text{for} \quad {i=\mathrm{p}}\\
    q_{\mathrm{e}}(p) + q_{\mathrm{e}^{\pm}}^{2 \mathrm{nd}}(p)\, &\text{for} \quad {i=\mathrm{e}}
    \end{cases}
    \label{eq:origTransportEQ}
\end{align}
where
$D(p) \propto p^\varkappa$ 
is the momentum diffusion coefficient, describing the interaction rate with the magnetic turbulence. Thus, the coronal condition (A1) is represented by both lines (\ref{eq:momDiffTerm}$+$\ref{eq:origTransportEQ}) of the previous equation. However, the first term (\ref{eq:momDiffTerm}) can be neglected for the starburst environment where SDA is typically subdominant as particles are more likely accelerated by supernova remnants (SNRs) via diffusive shock acceleration (DSA) which can be treated in a mathematically convenient way by a primary source rate $q_i(p)$. 
Therefore, we use the same transport equation with a two-zone approach as in the E+22 model for the starburst ring. 
Hence, we adopt 
\begin{equation}
    q_i(p) =
    \begin{cases}
    q_{0,i} \, (p/p_0)^s\, \exp(-p/\hat{p}_i) \, &\text{for the starburst}\\
    q_{0,i} \, \delta(p/p_{\rm inj}-1)  \, &\text{for the AGN corona,}
    \end{cases}
\end{equation}
where either already accelerated particles up to a maximal momentum $\hat{p}_i$ are injected into the starburst environment, or super-thermal particles enter the coronal environment with a distinct momentum $p_{\rm inj}$ specified by the velocity of the turbulent modes that subsequently accelerate these particles within the environment. Hereby, we adopt that the physical parameters in the vicinity of the SNRs are similar to what is found in the whole starburst environment, so that $\hat{p}_i$ is determined by the equilibrium of the characteristic acceleration and loss timescale in the starburst ring. In case of the AGN corona, we assume that a resonant interaction with the magnetosonic-type turbulence is only possible for suprathermal particles whose velocity is comparable to the velocity of the fast magnetosonic mode --- which is about the Alvf\'en speed $v_{\rm A}$ for low a plasma beta or magnetically dominated plasma. This condition sets the low-energy cutoff of the momentum diffusion coefficient and therefore, $p_{\rm inj}=p(v_{\rm A})$. 

The normalization $q_{0,{\rm p}}$ of the CR proton injection rate is determined as follows:
\begin{enumerate}
    \item[(i)] For the AGN corona we adopt that a certain fraction $f_{\rm gas}$ of the thermal gas pressure is transferred to CRs, i.e.\ $$f_{\rm gas}=\frac{c\int\text{d}p\,n_{\rm p}(p)\,p/3}{k_{\rm B}\,n_{\rm gas}\,T_{\rm gas}}\,,$$
where $T_{\rm gas}= m_{\rm p}c^2/(6r_{\rm c}k_{\rm B} )$ denotes the temperature of the protons and $k_{\rm B}$ is the Boltzmann constant.
\item[(ii)] For the starburst ring, a fraction $f_{\rm SN}$ of the total SN energy is transferred to CRs, so that
$$f_{\rm SN} = \frac{V_{\rm geo}^{\rm s}\,c\int\text{d}p\,q_{\rm p}(p)}{E_{\rm SN}\,\tau^{-1}_{\rm SN}}\,$$
where a typical total SN energy of $E_{\rm SN}=10^{51}\,\text{erg}$, as well as a SN rate of \citep{Veilleux+2005} $\tau^{-1}_{\rm SN}(\text{NGC\,1068})=0.51\,\text{yr}^{-1}$ and \citep{Laine+2006} $\tau^{-1}_{\rm SN}(\text{NGC\,7469})=0.33\,\text{yr}^{-1}$, respectively, is used. Moreover, the volume of the starburst ring of the considered sources are $V_{\rm geo}^{\rm s}(\text{NGC\,1068})=3.47\,\text{kpc}^3$ for an adopted outer radius of $R_{\rm out}^{\rm s}=1.5\,\text{kpc}$ and $V_{\rm geo}^{\rm s}(\text{NGC\,7469})=1.57\,\text{kpc}^3$ using $R_{\rm out}^{\rm s}=1\,\text{kpc}$, respectively. 
\end{enumerate}

Based on the resulting CR proton source rate $q_{\rm p}(p)$ and the associated CR proton number density $n_{\rm p}(p)$, the source rate of primary CR electrons (if present) is normalized due to the assumption of a quasi-neutral accelerator, i.e.\ $\int\text{d}p\,q_{\rm p}(p)\,\tau_{\rm esc,p}^{\rm acc}(p)=\int\text{d}p\,q_{\rm e}(p)\,\tau_{\rm esc,e}^{\rm acc}(p)$, in the case of the starburst ring; and a quasi-neutral environment, i.e.\ $\int\text{d}p\,n_{\rm p}(p)=\int\text{d}p\,n_{\rm e}(p)$, for the AGN corona. 
%The continuous losses, due to interaction with the ambient gas and photon targets, are incorporated in the $\langle\dot{p}\rangle<0$ parameter; and the catastrophic losses, which are important for the resulting steady-state solution even in the case of ineffective particle escape on a timescale $\tau_{\rm esc}^{\rm c}(p)\gg |p/\langle\dot{p}\rangle|$, are also taken into account --- in contrast to a previous treatment \citep{eichmann2023disentangling}. 
%All details on the energy losses, the escape and the acceleration timescales can be found in E+22. 

In addition, the term $q_{\rm e^{\pm}}^{\rm 2nd}(p)$ refers to the source rate of secondary electrons e$^-$ and positrons e$^+$, which is treated more accurately than in the E+22 model according to (A2). These secondaries can emerge either from hadronic processes directly or, for the AGN corona, from $\gamma\gamma$-pair attenuation. For the latter case, the $\gamma$-ray emission can also result from leptonic processes, such as e.g.\ inverse Compton (IC) scattering, leading to a non-linear dependency in the transport equation. To break this non-linearity we use an iterative approach, where leptonic $\gamma$-rays as well as the resulting pairs from $\gamma\gamma$-pair attenuation are calculated for each generation of the secondaries until the additional contribution becomes negligible. Finally, the hadronic contribution as well as the different leptonic contributions are summed up yielding the total source rate $q_{\rm e^{\pm}}^{\rm 2nd}(p)$.

%On the right panel of Fig.\ref{fig:had+lept} the solution of the transport equation (\ref{eq:transportEQ}) is shown for the best-fit scenario of the AGN corona of NGC 1068. In case of the non-thermal protons, the typical hard spectrum from SDA (see e.g.\ Ref.~\cite{walter2025stochastic} for more details) is obtained, with $n_{\rm p}^{\rm c}(E\lesssim 10^{13}\,\text{eV})\propto E^{1-\varkappa}$ at these energies where the acceleration is significantly faster than the energy losses ($\tau_{\rm acc}^{\rm c}\ll \tau_{\rm loss}^{\rm c}$), as well as the pile-up bump around the energy where those timescales are equal ($\tau_{\rm acc}^{\rm c}\sim \tau_{\rm loss}^{\rm c}$). The spectrum of non-thermal electrons is significantly softer, since primary electrons are not sufficiently accelerated by SDA ($\tau_{\rm acc}^{\rm c}\gg \tau_{\rm loss}^{\rm c}$), so that the overall spectrum consists only of the cooled secondaries.

%\begin{figure}[htbp]
%    \centering
%    \includegraphics[width=0.48\linewidth]{1.pdf}
%    \includegraphics[width=0.48\linewidth]{2.pdf}
%    \caption{Relativistic particles energy distribution in the AGN corona. On the left, the secondary electrons and positrons source rate. On the right, the differential energy density of the electrons (including the secondary contribution) and protons.} 
%    \label{fig:had+lept} 
%\end{figure}

\begin{figure*}
    \centering
    \includegraphics[width=0.48\linewidth]{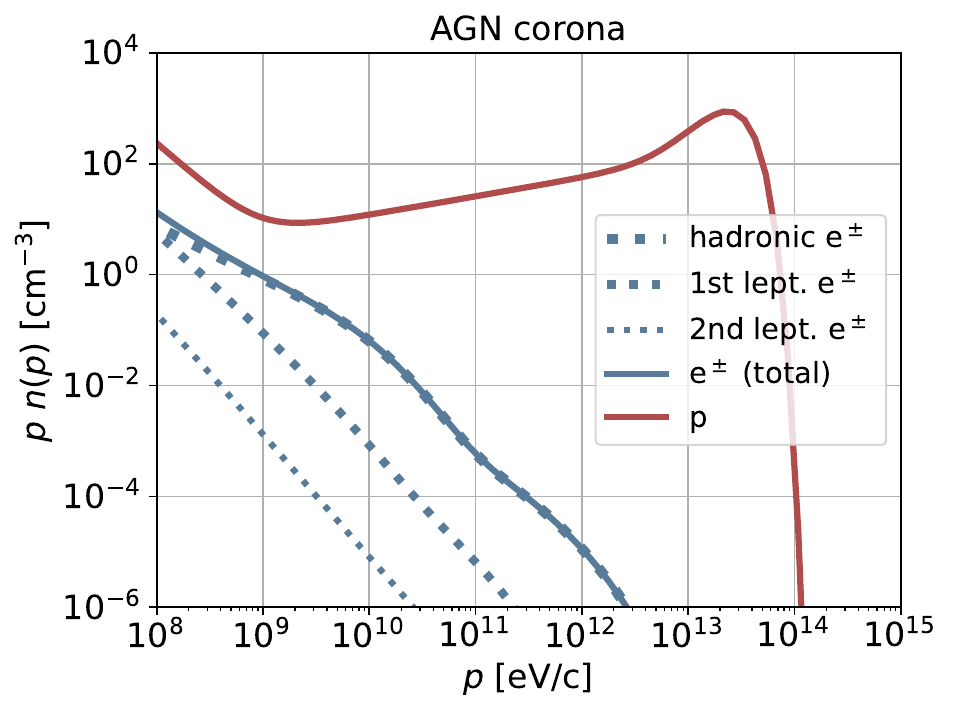}
    \includegraphics[width=0.48\linewidth]{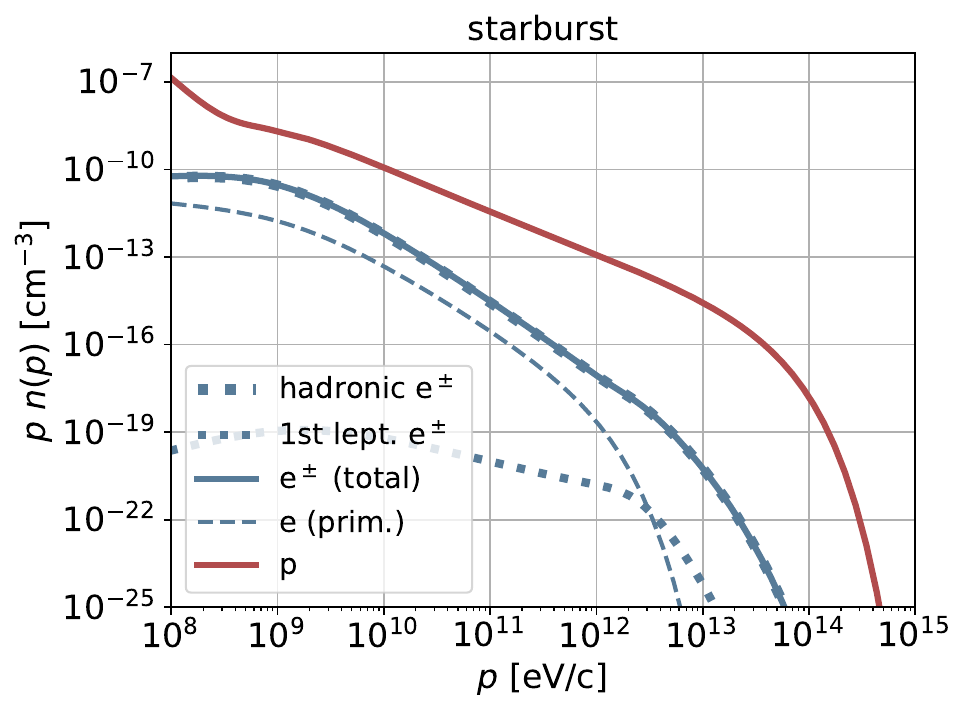}
    \caption{The best-fit model predictions of the CR electron (blue) and proton (red) distribution within the AGN corona (\emph{left}) and the starburst-ring (\emph{right}) of NGC\,1068.}
    \label{fig:crSED_NGC1068} 
\end{figure*}

The general solution of the steady-state momentum diffusion equation (\ref{eq:momDiffTerm}+\ref{eq:origTransportEQ}) has been provided by \cite{walter2025stochastic}. This leads to a significantly harder momentum distribution in the AGN corona than for the case of the simplified Eq.~(\ref{eq:origTransportEQ}) that has been used in E+22. Using the best-fit parameters (see Sect.~\ref{sec:res_NGC1068} for more details), the left panel in  Fig.~\ref{fig:crSED_NGC1068} illustrates these hard spectra for the CR protons (red line), whereas primary CR electrons do not reach such energies due to dominant Coulomb losses at low energies. Therefore, secondary electrons (blue solid line), which emerge predominantly from hadronic interaction processes, constitute the distribution function at high energies. Here, we do not account for stochastic acceleration which is typically inefficient due to strong synchrotron and IC cooling. For the starburst ring (right panel in Fig.~\ref{fig:crSED_NGC1068}), DSA is expected to yield a significantly softer injection spectra than SDA, that becomes even softer within the starburst ring in the case of CR electrons due to synchrotron and IC cooling. Also in this environment the high-energy electrons emerge --- for the considered best-fit scenario --- predominatly from hadronic interaction processes.

Based on the resulting CR distribution function $n_i(p)$, we can determine the resulting non-thermal photon and neutrino flux for the individual emission sites, such as in E+22. But in addition to the internal attenuation by the isotropic X-ray photon target within the AGN corona, we also account for the external attenuation (A3) as follows:

As shown in E+22 (and references therein) the internal attenuation can be associated to an attenuation factor $f_{\rm att}^{\rm int} = V_{\rm eff}(\tau)/V_{\rm geo}$, that is given by the ratio of the effective volume
\begin{equation}
    V_{\rm eff}^{\rm c}(\tau) = \frac{\pi\,R_{\rm c}^3}{\tau}\left[ 1 + \frac{(2\tau+1)\exp(-2\tau)-1}{2\tau^2}  \right]
\end{equation}
over the geometrical volume $V_{\rm geo}^{\rm c}=4\pi R_{\rm c}^3/3$, yielding $f_{\rm att}^{\rm int}\propto\tau^{-1}$ for an optical depth $\tau\gg 1$. Here, the coronal region is constrained to a radius $R_{\rm c}=r_{\rm c}\,\mathcal{R}_{\rm s}$, where $\mathcal{R}_{\rm s}=2GM/c^2$ denotes the Schwarzschild radius. 
Note that this relation only holds for a homogeneous spherical symmetric system with constant emission and absorption coefficients. 

However, this is no longer the case for the external attenuation, where the emission coefficient vanishes since gamma-rays are only produced inside of the AGN corona. Therefore, the external attenuation factor is provided by $f_{\rm att}^{\rm ext} = \exp(-\tau)$, where the optical depth $\tau$ accounts for the radially decreasing target photon density. 
%The external X-ray target photon number density with a frequency $\nu$ at a distance $s$ from the center of the corona is given by
The differential external X-ray target photon number $\diff N_{\rm X}$ per frequency $\diff\nu$, volume $\diff V$ and solid angle $d\Omega=2\pi\,d\mu$ at a distance $x$ from the center of the corona is given by
\begin{equation}
    n_{\rm X}^{\rm ext}(\nu,\,x,\,\mu) \equiv \frac{1}{2\pi}\,\frac{\diff N_{\rm X}}{\diff\nu\,\diff V\,\diff \mu'}\,\left| \frac{\diff \mu'}{\diff \mu} \right| = \frac{L_\nu}{2\pi\,R_{\rm c}^2c\,h\nu}\,\left( \frac{R_{\rm c}}{x} \right)^2\,\left| \frac{\diff \mu'}{\diff \mu} \right|
\end{equation}
where $\mu$ denotes the cosine of the interaction angle at a certain distance $x$ from the center of the system and $\mu'$ is at $x=0$ the cosine of the angle between the line of sight and the position on the surface of the corona, where the interacting X-ray escapes the system. 
The calculation of the Jacobian determinant yields the physically relevant solution\footnote{Note that the other mathematical solution, where the second term is subtracted, is not relevant for the considered scenario of photons that are isotropically leaving the coronal surface.}
\begin{equation}
    \left| \frac{\diff \mu'}{\diff \mu} \right| = \left|-\frac{2x}{R_{\rm c}}\mu  + \frac{\left(\frac{2}{R_{\rm c}}\right)^2\left[4\mu^3-2\mu\right]+2\mu}{2\sqrt{\left(\frac{x}{R_{\rm c}}\right)^2\left[\mu^4-\mu^2\right]+\mu^2}}\right|\,.
\end{equation}
Hereby, we adopt that the photon target is isotropic within the corona, i.e. no dependence on $\mu'$, and photons are leaving the surface isotropically, so that $n_{\rm X}^{\rm ext}$ depends on the cosine of the interaction angle $\mu$ (between the X-ray and the gamma-ray) at a distance $x>R_{\rm c}$. 
Using this photon target, the external optical depth due to $\gamma\gamma$-pair attenuation for a gamma-ray with an energy $E_\gamma$ is given by \citep{BreitWheeler1934, Heitler1954}
\begin{equation}\label{eq:alpha_gg}
    \tau^{\rm ext}_{\gamma\gamma}(E_\gamma) = 2\pi \int_{-1}^{1} \diff \mu\,(1-\mu) \int_{R_{\rm c}}^d \diff x\int_{\nu_{\rm th}}^{\infty} \diff \nu \, n_{\rm X}^{\rm ext}(\nu,x,\mu)\,\sigma_{\gamma \gamma}(E_\gamma, \nu, \mu)\,,
\end{equation}
with the pair production threshold frequency
\begin{equation}
    \nu_{\rm th} = \frac{2m_e^2c^4}{h\,E_\gamma(1-\mu)}\,,
\end{equation}
the cross section 
\begin{equation}
    \sigma_{\gamma \gamma}(E_\gamma, \nu, \mu)  = \frac{3\sigma_{\rm T}}{16}\left(1-\beta^2\right)\bigg[\left(3-\beta^4\right)\ln\Big(\frac{1+\beta}{1-\beta}\Big)-2\beta\left(2-\beta^2\right)\bigg]
\end{equation}
and the dimensionless center-of-momentum velocity
\begin{equation}
    \beta(E_\gamma, \nu, \mu) = \sqrt{1-\frac{2m_e^2c^4}{\nu h\,E_\gamma(1-\mu)}}\,.
\end{equation}
Moreover, we will also address a scenario,
where the gamma-ray emission region is completely confined within the corona at a radius $(1-r_{\rm sh})R_{\rm c}$, such that there is an outer shell containing only X-ray emitting coronal plasma, similar to what has previously been discussed by \cite{Inoue+2020}. In this case, the attenuation is also given by $f_{\rm att}^{\rm sh} = \exp(-\tau)$ due to the adopted absence of gamma-ray sources within the outer spherical shell between $(1-r_{\rm sh})R_{\rm c}$ and $R_{\rm c}$, where $r_{\rm sh}<1$ denotes the thickness of this shell in units of $R_{\rm c}$. But in contrast to the previous case of external attenuation, the target-photon distribution is still isotropic and constant within this shell leading to a much higher absorption probability. 

Such as in E+22, the spectral energy flux of the messenger type $m\in\{\gamma, \nu\}$, for a homogeneous source with constant emission coefficient $\epsilon_m$ and absorption coefficient $\alpha_m$ at a distance $d$ is given by \citep[e.g.][]{Gould1979}
\begin{equation}
    F_m(E_m) = \frac{E_m}{4\pi\,d^2}\,\epsilon_m(E_m)\,V_{\rm eff}(E_m)
\label{eq:fluxPred}
\end{equation}
with the effective emission volume
\begin{equation}
    V_{\rm eff}(E_m) = \int \diff^3 r\,\exp\left( - \alpha_m(E_m)\, |\vec{r}_{\rm s}-\vec{r}|\right)\,,
\end{equation}
where $\vec{r}_{\rm s}$ represents an arbitrary position on the surface of the emission volume (see E+22 for more details on $\epsilon_m$ and $\alpha_m$). Since the considered sources are at a distance $<100\,\text{Mpc}$ and $E_\gamma<10\,\text{TeV}$ the $\gamma$-attenuation by the extragalactic background light is negligible. 

%%%%%%%%%%%%%%%%%%%%%%%%%%%%%%%%%%%%%%%%%%%%%%%%%%%%%%%%%%%%%%%%%%%%%%%%%%%%%%%%%%%%%%%
%%%%%%%%%%%%%%%%%%%%%%%%%%%%%%%%%%%%%%%%%%%%%%%%%%%%%%%%%%%%%%%%%%%%%%%%%%%%%%%%%%%%%%%
%%%%%%%%%%%%%%%%%%%%%%%%%%%%%%%%%%%%%%%%%%%%%%%%%%%%%%%%%%%%%%%%%%%%%%%%%%%%%%%%%%%%%%%
\section{Methods}\label{sec:methods}
In the following, we apply our model to explain the non-thermal observational data from NGC\,1068 and NGC\,7469. However, the non-thermal phenomena that can be attributed to their starburst ring or the AGN corona only emerge at the radio\footnote{Note that at radio/mm wavelengths, the emission can be a combination of thermal and non-thermal radiation.}- and gamma-ray (only for NGC\,1068) band as well as at about TeV-PeV energies for the neutrinos. Hence, this lack of fit data demands a consideration of prior information on the model parameters to avoid overfitting. 
%For the AGN corona we are mostly limited to indirect information that result from simulations.   
For the majority of parameters, a truncated normal distribution (mostly in log-space) is applied, where minimal or maximal values represent limits that result either from fundamental physical constraints or observational certainty. The standard deviation $\sigma$, describes the adopted confidence on the expectation $\mu$. Only for $r_{\rm c}$, there is no expectation (but a clear minimal and maximal value), so that a uniform distribution is used (see Table \ref{tab:priors}).
\begin{table*}[htbp]
\centering
\begin{threeparttable}
\caption{Prior distributions $\Pi(\theta)$ used for model parameters $\theta$ of starburst ring (str) and AGN corona (cor).}
\label{tab:priors}
\begin{tabular}{lllll}
\toprule
\textbf{Parameter} ($\theta$) & \textbf{Env.} & \textbf{Prior parameters} & \textbf{Note} \\
\midrule
$\ln(\eta_{\rm c}^{-1})$ & cor &  $\mu = \ln(0.1)$, $\sigma = \ln(10)$, $\ln(\eta_{\rm c}^{-1})<\ln(0.9)$ & turbulence strength\tnote{a)} \\
$\ln(f_{\rm gas})$ & cor &  $\mu = \ln(0.1)$, $\sigma = \ln(100)$, $\ln(f_{\rm gas})<\ln(1)$ & gas-to-CR pressure ratio\tnote{b)} \\
$\ln(\beta_{\rm B})$ & cor &  $\mu = \ln(0.5)$, $\sigma = \ln(10)$, $\ln(0.05)<\ln(\beta_{\rm B})<\ln(5)$ & plasma beta\tnote{c)} \\
$\ln(\omega_{\rm T})$ & cor &  $\mu = \ln(0.5)$, $\sigma = \ln(10)$, $\ln(0.05)<\ln(\omega_{\rm T})<\ln(5)$ & optical Thomson depth\tnote{d)} \\
$r_{\rm c}$ & cor &  $r_{\rm c}\in [1, 1000]$ & coronal radius  \\
%%%
$\ln(\eta_{\rm s}^{-1})$ & str &  $\mu = \ln(0.1)$, $\sigma = \ln(10)$, $\ln(\eta_{\rm s}^{-1})<\ln(0.9)$ & turbulence strength\tnote{a)} \\
$\ln(f_{\rm SN})$ & str &  $\mu = \ln(0.1)$, $\sigma = \ln(100)$, $\ln(f_{\rm SN})<\ln(0.5)$ & SN-to-CR energy ratio\tnote{e)} \\
$\ln(B_{\rm s}/1\mu\text{G})$ & str &  $\mu = \ln(100)$, $\sigma = \ln(3)$, $\ln(10)<\ln(B_{\rm s}/1\mu\text{G})<\ln(1000)$ & av.\ magnetic field strength\tnote{f)} \\
$\ln(n_{\rm gas}/1\text{cm}^{-3})$ & str &  $\mu = \ln(500)$, $\sigma = \ln(5)$, $\ln(50)<\ln(n_{\rm gas}/1\text{cm}^{-3})<\ln(5000)$ & av.\ gas density\tnote{g)} \\
$s$ & str & $\mu = -1$, $\sigma = 0.5$, $0.5>s>-2.5$ & source rate spectral index\tnote{h)} \\
\bottomrule
\end{tabular}
More details on the motivation for the parameter expectation can be found here:
\begin{tablenotes}
\item[a)] E.g.~\cite{Fiorillo+2024b} for the AGN corona and \cite{Beck2012} for the starburst. In principle also larger $\mu$ values can be expected, that however, leads to inconsistency with the adopted quasilinear approximation in Eq.~(\ref{eq:momDiffTerm}+\ref{eq:origTransportEQ}).
\item[b)] \cite{PetrosianLiu2004}.
\item[c)] \cite{Jiang+2019,Jiang+2014,MillerStone2000}.
\item[d)] \cite{Ricci+2018,MerloniFabian2001}.
\item[e)] E.g.~\cite{Haggerty+2020}, where the normalized CR pressure
quickly converges towards about $10\%$. Additional uncertainty results from the adopted typical SN energy of $10^{51}\,\text{erg}$.
\item[f)] E.g.\ \cite{Thompson+2006}.
\item[g)] E.g.\ \cite{Hsieh+2011,garcia2014molecular,Izumi+2015}.
\item[h)] E.g.~\cite{drury1983introduction} for an ideal, non-relativistic, strong shock with Bohm-diffusion ($\tau_{\rm esc}^{\rm acc}\propto p^{-1}$) in the vicinity of the accelerator, so that $q_i=n_i^{\rm acc}/\tau_{\rm esc}^{\rm acc}\propto p^{-1}$ is expected.
\end{tablenotes}
\end{threeparttable}
\end{table*}

To describe the $\gamma$-ray flux from NGC\,1068 and NGC\,7469, we need to account for the coronal target photon field, which can be estimated from the characteristic X-ray flux at these energies. In the case of NGC\,7469, the impact of the torus attenuation is assumed to be negligible and the spectral distribution from \cite{mehdipour2018multi} with a coronal X-ray luminosity (between 2 and 10 keV) of $L_X = 1.7 \times 10^{43}\,\text{erg}/\text{s}$ is used to model the intrinsic target field. Hereby, the \textit{Chandra}-HETGS, \textit{HST}-COS and \textit{Swift}-UVOT data from 2015 have been fitted yielding an X-ray flux that is just slightly larger than the results from previous epochs.

However, for the Seyfert 2 galaxy NGC\,1068 a more detailed modeling of the impact of the torus is necessary to obtain the intrinsic coronal target photon field. NuSTAR and XMM-Newton monitoring campaigns \citep{Marinucci+2016} yield an intrinsic luminosity of $L_{\rm X}=7^{+7}_{-4}\times 10^{43}\,\text{erg/s}$ between $(2-10)\,\text{keV}$, which is used in the following. 
Consequently, the black hole mass of NGC\,1068 is estimated to \citep[e.g.][]{Mayers+2018_arxiv,GRAVITY2020} $M_{\rm BH}=1.7\times 10^7\,M_\odot$, which leads to an Eddington luminosity of  $L_{\rm Edd}=2.1\times 10^{45}\,\text{erg s}^{-1}$. Moreover, the empirical relation of \cite{Hopkins+2007} is used to determine $L_{\rm bol}$ based on $L_{\rm X}$. Finally, the resulting Eddington ratio, $\lambda_{\rm Edd} = L_{\rm bol}/L_{\rm Edd}$, is used to model the spectral behavior of the X-ray corona field by a power-law, with a photon index $\Gamma_X\simeq 2+0.167\,\log(\lambda_{\rm Edd})$ (see \cite{Trakhtenbrot+2017}) and an exponential cutoff at \citep{Ricci+2018} $E_{\rm X,cut} \simeq [-74 \log(\lambda_{\rm Edd}) + 1.5\times 10^2 ]\,\text{keV}$. Moreover, the parametrized model by \cite{Ho2008} is used between $2\,\text{eV}$ and $2\,\text{keV}$ for the thermal UV photon spectrum from the disk.

In addition to these coronal X-ray fields, the AGNs present extended diffuse emission produced on significantly larger scales, whose physical origin remains ambiguous. This extended emission can be dominant within the radio and IR bands (depending on the angular resolution of the observations). Here we stay agnostic to this extended component and use the approach from \cite{mutie2025consistent} where, in addition to the non-thermal coronal contribution, three components from the vicinity of the AGN are taken into account:
\begin{enumerate}
    \item[(i)] An extended synchrotron contribution by diffuse CR electrons, that may be related to the jet, which emits a spectrum according to
    \begin{equation}
        S_{\rm sy}(\nu) = A_{\rm sy}\,\left( \frac{\nu}{\nu_0} \right)^{\alpha_{\rm sy}}\,.
    \end{equation}
    \item[(ii)] Free--free radiation by the ionized gas, that extends beyond the spatial scales of the X-ray corona, whose optical thin spectrum is given by
    \begin{equation}
        S_{\rm ff}(\nu) = A_{\rm ff}\,\left( \frac{\nu}{\nu_0} \right)^{\alpha_{\rm ff}}\,.
    \end{equation}
    \item[(iii)] Thermal emission from the dust particles within the unresolved core (i.e. at a radial distance $\lesssim 1\,\text{pc}$ and $\lesssim 100\,\text{pc}$ from the core of NGC\,1068 and NGC\,7469, respectively) that can be modeled as a modified-blackbody spectrum yielding
    \begin{equation}
        S_{\rm d}(\nu) = A_{\rm d}\,\left( \frac{\nu}{\nu_0} \right)^{2}\,(1-\exp[-(\nu/\nu_{\tau_{1}})^{\beta_{\rm d}}])\,,
    \end{equation}
\end{enumerate}
where $\nu_{\tau_1} \sim 800$~GHz is the frequency at which the dust becomes optically thick and $\beta_\mathrm{d}\sim 1.5$--2 is related to the dust opacity law. 
%For NGC\,1068, \cite{mutie2025consistent} showed that component (i) has a minor impact on the explanation of the so-called mm-bump, which is a spectral feature that is typically associated with the emission from the compact corona. Therefore, only free--free radiation (as the most dominant contributor at these frequencies) and thermal dust emission that becomes in particular relevant towards higher frequencies. 
In general, the diffuse synchrotron component (i) is dominant at low frequencies, then free--free becomes more relevant, and at high frequencies (submm to FIR regime) the dust dominates. In some sources --including NGC\,1068 and NGC\,7469-- the SED shows an additional bump around mm wavelengths that is associated with emission from the compact corona.
For NGC\,1068, \cite{mutie2025consistent} showed that component (i) has a negligible impact on the explanation of this mm-bump. Therefore, only free--free and thermal dust emission is taken into account. 
For NGC\,7469 a similar MCMC analysis has been performed (del Palacio et al., in prep.), yielding an extended synchrotron contribution that becomes relevant at frequencies $\lesssim 100\,\text{GHz}$, so this component is also taken into account. All details on the adopted parameters for these fixed  contributions from the vicinity of the AGN are given in Table \ref{tab:fixedParamRadio}.
\begin{table}[h!]
\centering
\caption{Fixed Radio Parameters for NGC\,1068 and NGC\,7469}
\label{tab:fixedParamRadio}
\begin{tabular}{lcc}
\toprule
\textbf{Parameter} & \textbf{NGC\,1068} & \textbf{NGC\,7469} \\
\midrule
$\nu_{0}$ [GHz] & 100 & 100 \\
$A_{\rm sy}$ [Jy] & --- & 0.50 \\
$\alpha_{\rm sy}$ & --- & $-$1.33 \\
$A_{\rm ff}$ [Jy] & 10.47 & 1.34 \\
$\alpha_{\rm ff}$ & $-$0.1 & $-$0.1 \\
$A_{\rm d}$ [Jy] & 0.022 & 1.585 \\
$\nu_{\tau_{1}}$ [GHz] & 800 & 820 \\
$\beta_{\rm d}$ & 2 & 1.78 \\
\bottomrule
\end{tabular}
\end{table}

\begin{table}[h!]
\centering
\begin{threeparttable}
\caption{Fit Data for NGC\,1068.}
\label{tab:fitData1068}
\begin{tabular}{lcc}
\toprule
\textbf{Env.} & \textbf{Energy} [eV] & \textbf{Flux} [eV\,s$^{-1}$\,cm$^{-2}$] \\
\midrule
str\tnote{a)} & $0.6\times 10^{-5}$ & $(7\pm 2)\times 10^{-3}$ \\
str\tnote{a)} & $2.1\times 10^{-5}$ & $(10\pm 3)\times 10^{-3}$ \\
str\tnote{a)} & $6.2\times 10^{-5}$ & $(10\pm 3)\times 10^{-3}$ \\
cor\tnote{b)} & $0.78\times 10^{-4}$ & $(0.153\pm 0.001)\times 10^{-2}$  \\
cor\tnote{b)} & $0.80\times 10^{-4}$ & $(0.153\pm 0.001)\times 10^{-2}$  \\
cor\tnote{b)} & $0.82\times 10^{-4}$ & $(0.155\pm 0.001)\times 10^{-2}$  \\
cor\tnote{b)} & $0.91\times 10^{-4}$ & $(0.165\pm 0.001)\times 10^{-2}$  \\
cor\tnote{b)} & $0.92\times 10^{-4}$ & $(0.174\pm 0.001)\times 10^{-2}$  \\
cor\tnote{b)} & $0.95\times 10^{-4}$ & $(0.172\pm 0.001)\times 10^{-2}$  \\
cor\tnote{b)} & $3.89\times 10^{-4}$ & $(0.645\pm 0.035)\times 10^{-2}$  \\
cor\tnote{b)} & $4.14\times 10^{-4}$ & $(0.637\pm 0.031)\times 10^{-2}$  \\
cor\tnote{b)} & $9.97\times 10^{-4}$ & $(1.59\pm 0.09)\times 10^{-2}$  \\
cor\tnote{b)} & $10.6\times 10^{-4}$ & $(1.58\pm 0.01)\times 10^{-2}$  \\
cor\tnote{b)} & $14.3\times 10^{-4}$ & $(3.47\pm 0.06)\times 10^{-2}$  \\
cor\tnote{b)} & $14.8\times 10^{-4}$ & $(3.10\pm 0.02)\times 10^{-2}$  \\
cor\tnote{b)} & $19.7\times 10^{-4}$ & $(5.57\pm 0.15)\times 10^{-2}$  \\
cor\tnote{b)} & $20.0\times 10^{-4}$ & $(5.91\pm 0.12)\times 10^{-2}$  \\
cor\tnote{b)}\tnote{b)} & $28.5\times 10^{-4}$ & $(8.6\pm 1.1)\times 10^{-2}$  \\
cor\tnote{b)} & $28.8\times 10^{-4}$ & $(7.40\pm 0.74)\times 10^{-2}$  \\
cor\tnote{b)} & $29.2\times 10^{-4}$ & $(7.27\pm 0.84)\times 10^{-2}$  \\
str+cor\tnote{c)} & $0.18\times 10^{9}$ & $0.85 \pm 0.18$ \\
str+cor\tnote{c)} & $0.56\times 10^{9}$ & $0.81\pm 0.10$ \\
str+cor\tnote{c)} & $1.7\times 10^{9}$ & $0.61^{+0.08}_{-0.07}$ \\
str+cor\tnote{c)} & $5.5\times 10^{9}$ & $0.30^{+0.08}_{-0.07}$ \\
str+cor\tnote{c)} & $17\times 10^{9}$ & $0.42^{+0.14}_{-0.11}$ \\
str+cor\tnote{c)} & $0.5\times 10^{11}$ & $<0.18$ \\
str+cor\tnote{c)} & $1.7\times 10^{11}$ & $<0.39$ \\
str+cor\tnote{d)} & $1.8\times 10^{11}$ & $<0.69$ \\
str+cor\tnote{d)} & $5.6\times 10^{11}$ & $<0.17$ \\
str+cor\tnote{d)} & $18\times 10^{11}$ & $<0.07$ \\
str+cor\tnote{d)} & $56\times 10^{11}$ & $<0.18$ \\
\midrule
str+cor\tnote{e)} & $1.4\times 10^{12}$ & $96^{+60}_{-65}$ \\
str+cor\tnote{e)} & $3.1\times 10^{12}$ & $38^{+24}_{-22}$ \\
str+cor\tnote{e)} & $14\times 10^{12}$ & $6^{+8}_{-5}$ \\
\bottomrule
\end{tabular}
Due to angular resolution limitations in some cases the contributions from the starburst ring (str) and the AGN corona (cor) are disentangled. Messenger type of the data and associated reference are as follows:
\begin{tablenotes}
\item[a)] $\gamma$: New data --- see Sect.~\ref{sec:newData} for details. %\cite{Wynn-Williams+1985}
\item[b)] $\gamma$: \cite{mutie2025consistent}.
\item[c)] $\gamma$: \change{Using a custom time coverage of J0242.6-0000 between 2008 and 2020 in the energy range
$100\,\text{MeV}-300\,\text{GeV}$ taken from E+22. More details on the analysis can be found in \cite{Kun+2022} and \cite{Kun+2023}.}
\item[d)] $\gamma$: \cite{Acciari+2019}.
\item[e)] $\nu$: Using three characteristic energies within the 68\% confidence interval as reported by \cite{IceCube2022_Sci}.
\end{tablenotes}
\end{threeparttable}
\end{table}

\begin{table}[h!]
\centering
\begin{threeparttable}
\caption{Fit Data for NGC\,7469.}
\label{tab:fitData7469}
\begin{tabular}{lcc}
\toprule
\textbf{Env.} & \textbf{Energy} [eV] & \textbf{Flux} [eV\,s$^{-1}$\,cm$^{-2}$] \\
\midrule
str+cor\tnote{a)} & $0.6\times 10^{-5}$ & $(1.68\pm 0.07)\times 10^{-3}$  \\
str+cor\tnote{b)} & $2.0\times 10^{-5}$ & $(2.1\pm 0.2)\times 10^{-3}$  \\
str+cor\tnote{b)} & $3.5\times 10^{-5}$ & $(2.63\pm 0.03)\times 10^{-3}$  \\
str+cor\tnote{b)} & $9.3\times 10^{-5}$ & $(2.46\pm 0.07)\times 10^{-3}$  \\
str+cor\tnote{c)} & $42\times 10^{-5}$ & $(5.0\pm 0.5)\times 10^{-3}$  \\
str+cor\tnote{c)} & $46\times 10^{-5}$ & $(5.9\pm 0.5)\times 10^{-3}$  \\
cor\tnote{c)} & $2\times 10^{-5}$ & $(0.84\pm 0.08)\times 10^{-3}$ \\
cor\tnote{c)} & $4\times 10^{-5}$ & $(0.79\pm 0.08)\times 10^{-3}$ \\
cor\tnote{c)} & $6\times 10^{-5}$ & $(0.65\pm 0.06)\times 10^{-3}$ \\
cor\tnote{c)} & $14\times 10^{-5}$ & $(0.69\pm 0.07)\times 10^{-3}$ \\
cor\tnote{c)} & $37\times 10^{-5}$ & $(1.8\pm 0.2)\times 10^{-3}$ \\
cor\tnote{c)} & $41.7\times 10^{-5}$ & $(1.8\pm 0.2)\times 10^{-3}$ \\
cor\tnote{c)} & $42.4\times 10^{-5}$ & $(1.7\pm 0.2)\times 10^{-3}$ \\
cor\tnote{c)} & $46\times 10^{-5}$ & $(1.9\pm 0.4)\times 10^{-3}$ \\
cor\tnote{c)} & $89\times 10^{-5}$ & $(5.0\pm 0.5)\times 10^{-3}$ \\
cor\tnote{c)} & $95\times 10^{-5}$ & $(5.0\pm 0.5)\times 10^{-3}$ \\
cor\tnote{c)} & $195\times 10^{-5}$ & $(37\pm 20)\times 10^{-3}$ \\
cor\tnote{c)} & $198\times 10^{-5}$ & $(42\pm 7)\times 10^{-3}$ \\
cor\tnote{c)} & $200\times 10^{-5}$ & $(41\pm 20)\times 10^{-3}$ \\
str+cor\tnote{d)} &  $0.2\times 10^{9}$ & $<0.02$  \\
str+cor\tnote{d)} &  $0.6\times 10^{9}$ & $<0.02$  \\
str+cor\tnote{d)} &  $1.8\times 10^{9}$ & $<0.03$  \\
str+cor\tnote{d)} &  $5.6\times 10^{9}$ & $<0.06$  \\
str+cor\tnote{d)} &  $17.8\times 10^{9}$ & $<0.04$  \\
str+cor\tnote{d)} &  $56.2\times 10^{9}$ & $<0.13$  \\
str+cor\tnote{d)} &  $178\times 10^{9}$ & $<0.39$  \\
str+cor\tnote{d)} &  $562\times 10^{9}$ & $<1.75$  \\
\midrule
%str+cor, $\nu$\tnote{e)} & $1.61\times 10^{14}$ & $(32\pm 16)$ \\  %maybe not needed?!
%str+cor, $\nu$\tnote{f)} & $1.61\times 10^{14}$ & $(7.3\pm 4.2)$ \\ %maybe not needed?!
str+cor\tnote{e)} & $1.61\times 10^{14}$ & $(2.4\pm 1.4)$ \\
\bottomrule
\end{tabular}
Due to angular resolution limitations in some cases the contributions from the starburst ring (str) and the AGN corona (cor) are disentangled. Messenger type of the data and associated reference are as follows:
\begin{tablenotes}
\item[a)] $\gamma$: New data --- see Sect.~\ref{sec:newData} for details.%VLA L-Band (1.5GHz) A-array data from Project ID: 23A-324 obtained through the NRAO archive. 
\item[b)] $\gamma$: Taken from \cite{Clemens+08}. %Clemens et al (2008): https://www.aanda.org/articles/aa/pdf/2008/49/aa7224-07.pdf
\item[c)] $\gamma$: mm data retrieved from the ALMA archive (\url{https://almascience.nrao.edu/aq/}).
\item[d)] $\gamma$: New data --- see Sect.~\ref{sec:newData} for details.%\textit{Fermi}-LAT $95$\% CL upper limits of $\approx 15.5$\,years of LAT data (\url{https://fermi.gsfc.nasa.gov/cgi-bin/ssc/LAT/LATDataQuery.cgi}) (from 2008-08-04 to 2024-01-01) derived using the \texttt{sed()} method of \textsc{Fermipy} \citep{2017ICRC...35..824W}. % Fermi data (provided by Patrik) 
%\item[e)] Based on the 90\% confidence interval $(1.76,\,19.6)\,\text{eV}\,\text{cm}^{-2}\,\text{s}^{-1}$ using 4 years of data and associating two muon neutrinos \citep{sommani2025two}. % taken from Sommani et al. (2025) https://arxiv.org/pdf/2403.03752
%\item[f)] Based on the 90\% confidence interval $(0.137,\,4.74)\,\text{eV}\,\text{cm}^{-2}\,\text{s}^{-1}$ using 12 years of data and associating  one neutrino \citep{sommani2025two}. % taken from Sommani et al. (2025) https://arxiv.org/pdf/2403.03752
%\item[g)] 
\item[e)] $\nu$: Based on the average energy of the two neutrinos in the 90\% confidence energy interval with an 68\% confidence flux range of $(0.007,\,1.58)\,\text{eV}\,\text{cm}^{-2}\,\text{s}^{-1}$ from \cite{IceCube2025_arxiv}. % taken from Sommani et al. (2025) https://arxiv.org/pdf/2403.03752
\end{tablenotes}
\end{threeparttable}
\end{table}

Under consideration of this fixed components we fit the extended and compact radio emission, as well as the $\gamma$-ray and neutrino data (in case of NGC\,7469 only $\gamma$-ray upper limits are available) from these two Seyfert-Starburst-composite galaxies (see Table~\ref{tab:fitData1068} and \ref{tab:fitData7469}). Hereby, the standard sampler for Markov Chain Monte Carlo \texttt{EMCEE} \citep{Foreman-Mackey+2013} is used, where for a given set of model parameters ($\theta$) under consideration of the previously mentioned radio and high-energy data ($D$), the posterior probability density function $P(\theta|D)\propto \exp[-\chi^2(\theta)/2]\,\Pi(\theta)$ is computed. Here, $\Pi(\theta)$ denotes the priors adopted for each parameter (see Table~\ref{tab:priors} for more details), and the likelihood depends on $\chi^2(\theta) = \sum_{i}(F_{i}(\theta)-F_{{\rm D},i})^2/(\Delta F_{{\rm D},i})^2$, where $F_{i}(\theta)$ refers to the predicted flux (\ref{eq:fluxPred}) and $F_{{\rm D},i}$ and $\Delta F_{{\rm D},i}$ denote the observed flux and its uncertainty, respectively, for a certain messenger at a certain energy ($i$). Note, that the relative uncertainty $\Delta F_{{\rm D},i}/F_{{\rm D},i}$ of some of the ALMA data (cor) is multiple orders of magnitudes smaller than what is given at high energies, so that the likelihood would become completely dominated by these data points. In order to avoid such behavior and to address the additional uncertainty that results from the limited spatial resolution as well as the limited accuracy of our model prediction, we use in the following a minimal uncertainty of $\Delta F_{{\rm D},i}\geq 0.15\,F_{{\rm D},i}$ unless otherwise stated. In case of NGC\,7469 the non-thermal flux prediction of the starburst ring does not depend significantly on the coronal contribution, if we adopt that the high-energy neutrinos need to emerge from the AGN corona. Hence, we perform two independent MCMC fits for these two environments of NGC\,7469 due to computational efficiency. The fixed extended AGN emission of NGC\,7469 can become dominant at $1.4\,\text{GHz}$, if free-free absorption is negligible at these frequencies. However, the free-free optical depth $\tau_{\rm ff}(1.4\,\text{GHz})$ strongly depends on the details of the gas distribution within this extended region. Therefore, the given flux at this frequency rather serves as an upper limit for the flux contribution of the starburst ring. Hence, we artificially increase the relative inferior data uncertainty of the low resolution flux at $1.4\,\text{GHz}$ to $100\%$ in the likelihood of the MCMC.
%%%%%%%%%%%%%%%%%%%%%%%%%%%%%%%%%%%%%%%%%%%%%%%%%%%
\subsection{New Data Acquisition}
\label{sec:newData}
Most of the data that is used in the MCMC fit has already been published as indicated in Table \ref{tab:fitData1068} and \ref{tab:fitData7469}. But in the case of NGC\,1068 we reinvestigated the large scale radio emission from the starburst ring. Due to the non-vanishing radio contribution by the jet of NGC\,1068, which is not taken into account in this work, we need to ensure that the chosen spatial scales of the large-scale radio data exclude the jet contribution. Therefore, we use the publicly available map from the FIRST survey \citep{White1997} at $1.43\,\text{GHz}$, which provides a total flux from the whole $\sim 25^{\prime\prime}$ aperture of $4.41\pm0.01\,\text{Jy}$. Guided by a higher resolution map at $4.8\,\text{GHz}$ we identified the jet contribution, and consequently removed the flux coming from an ellipse of $\sim 13^{\prime\prime}\times 9^{\prime\prime}$, leading to $1.43\,\text{GHz}$ flux by the starburst ring of $0.3\pm0.1\,\text{Jy}$. Hereby, we used a conservative estimate on the uncertainty that includes the uncertainty of the chosen size. Hence, about 7\% of the total flux at $1.43\,\text{GHz}$ emerges from the starburst ring, which is about a factor of three less than what has been used in E+22 but in good agreement with a previous analysis by \cite{Mutie2019}. Using the large-scale spectral behavior from \cite{Wynn-Williams+1985} with $S_\nu\propto \nu^{\alpha_\nu}$, where $\alpha_\nu=-0.8\,(-1)$ between 20 and 6 cm (between 6 and 2 cm), we estimated the flux at $5\,\text{GHz}$ and $15\,\text{GHz}$ --- see Table \ref{tab:fitData1068}. Hereby, we adopted a spectral uncertainty of $\Delta\alpha_\nu= 0.05$.

In the case of NGC\,7469, there is no significant jet contribution, so that all components of the large-scale radio emission are considered in our model. Using the VLA L-Band A-array data from Project ID: 23A-324 obtained through the NRAO archive, we obtain a total flux at $1.5\,\text{GHz}$ of $0.18\pm0.00054\,\text{Jy}$ from an aperture of $\sim 1.43^{\prime\prime} \times 0.82^{\prime\prime}$. The flux was estimated using a single-component Gaussian on the source, with the uncertainty of the flux estimated using the rms noise. %\textbf{[@Crystal: Add the missing information; and potentially more if needed] Information addeed, I commented this out - Crystal}. 

At higher energies, we use $\sim15.5$\,years of \textit{Fermi}-LAT \citep{2009ApJ...697.1071A} data\footnote{\url{https://fermi.gsfc.nasa.gov/cgi-bin/ssc/LAT/LATDataQuery.cgi}} (from 2008-08-04 to 2024-01-01) between $100$\,MeV and $1$\,TeV. We selected a region of interest (ROI) of $15^\circ$ centered on the optical position of NGC~7469 \change{and Pass 8 SOURCE class events (evclass = 128)}. \change{A zenith angle cut of $90^{\circ}$ was applied.} The analysis was performed using the P8R3 SOURCE V3 instrument response functions \citep{2018arXiv181011394B}, together with the most recent Galactic (\texttt{gll\_iem\_v07}) and isotropic diffuse emission (\texttt{iso\_P8R3\_SOURCE\_V3\_v1}) models, and the \textit{Fermi}-LAT 4FGL DR4 source catalog \citep{4fgl}. The LAT data were reduced with \textsc{Fermipy} \citep{2017ICRC...35..824W}, following the standard analysis procedure. This included optimizing the ROI, freeing \change{and fitting the normalizations} of sources within $2^\circ$ of the target, and performing a global likelihood fit. We subsequently added a trial point source at the position of the Seyfert galaxy and repeated the likelihood analysis. To assess the presence of $\gamma$-ray emission from NGC~7469 over the full 15.5-yr \textit{Fermi}-LAT dataset, we generated test statistic (TS) maps and evaluated the TS value at the source position. We obtained a TS value consistent with zero, indicating that NGC~7469 is not detected by \textit{Fermi}-LAT. This result is consistent with the absence of the source in the FGL catalogs, and rules out even a marginal detection. Hereby, the \texttt{sed()} method of \textsc{Fermipy} is applied to determine $95$\% CL \change{spectral} upper limits --- see Table \ref{tab:fitData7469}.
%%%%%%%%%%%%%%%%%%%%%%%%%%%%%%%%%%%%%%%%%%%%%%%%%%%%%%%%%%%%%%%%%
%%%%%%%%%%%%%%%%%%%%%%%%%%%%%%%%%%%%%%%%%%%%%%%%%%%%%%%%%%%%%%%%%%%%%%%%%%%%%%%%%%%%%%%
%%%%%%%%%%%%%%%%%%%%%%%%%%%%%%%%%%%%%%%%%%%%%%%%%%%%%%%%%%%%%%%%%%%%%%%%%%%%%%%%%%%%%%%
\section{Results}\label{sec:results}
In general we noticed that the coronal radio prediction is strongly constraint by the given mm-data, despite of the additional extended diffuse component which is fixed from the given data at few $\times 10\,\text{GHz}$. However, this still leaves a huge variety of scenarios for the corresponding high-energy neutrino flux (bright red transparent lines in the right plot of Fig.~\ref{fig:SED_NGC1068} and \ref{fig:SED_NGC7469}), that all face some issues as discussed in more detail in the following.
\subsection{NGC\,1068}
\label{sec:res_NGC1068}
\begin{figure*}
    \centering
    \includegraphics[width=0.48\linewidth]{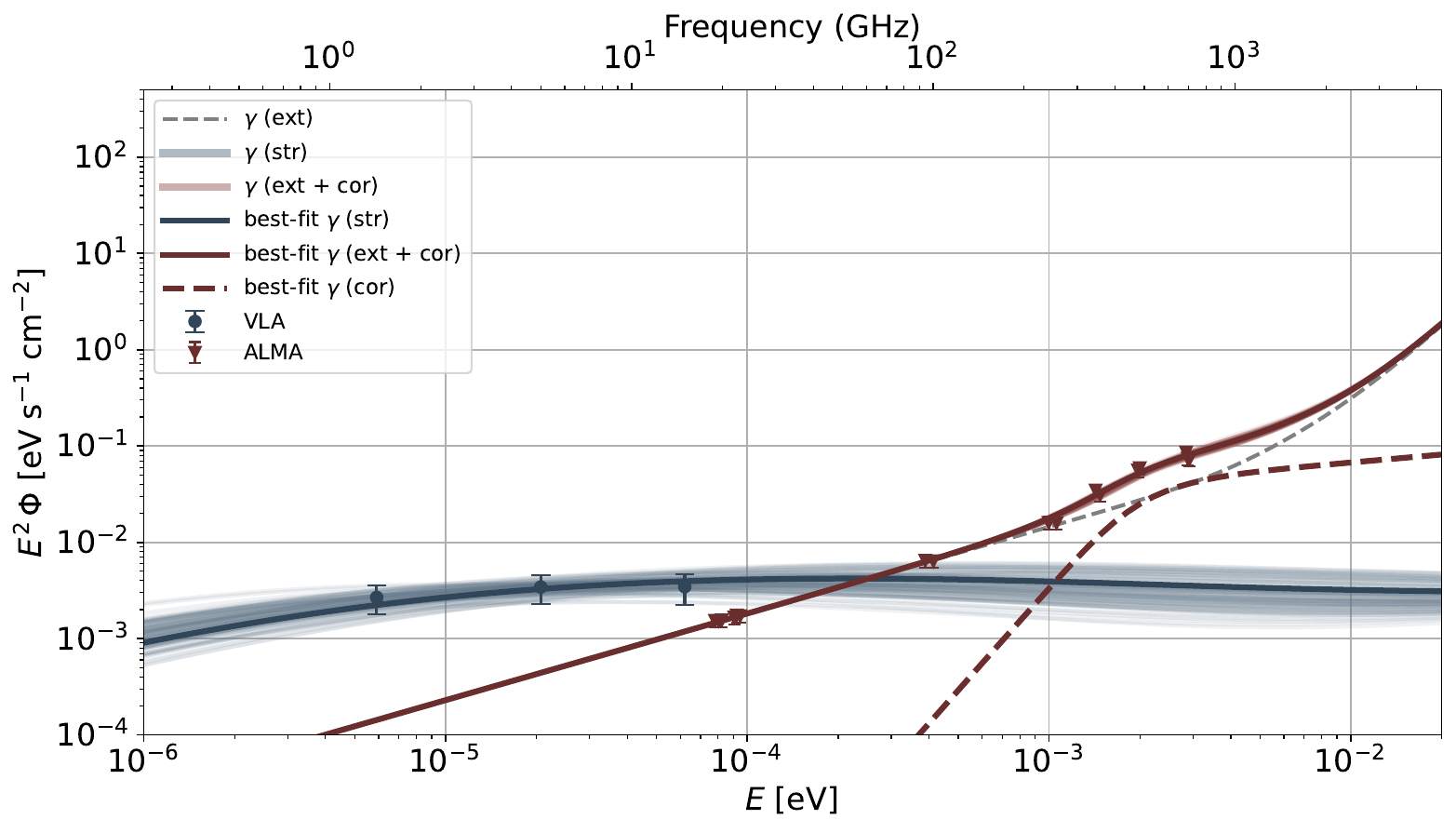}
    \includegraphics[width=0.48\linewidth]{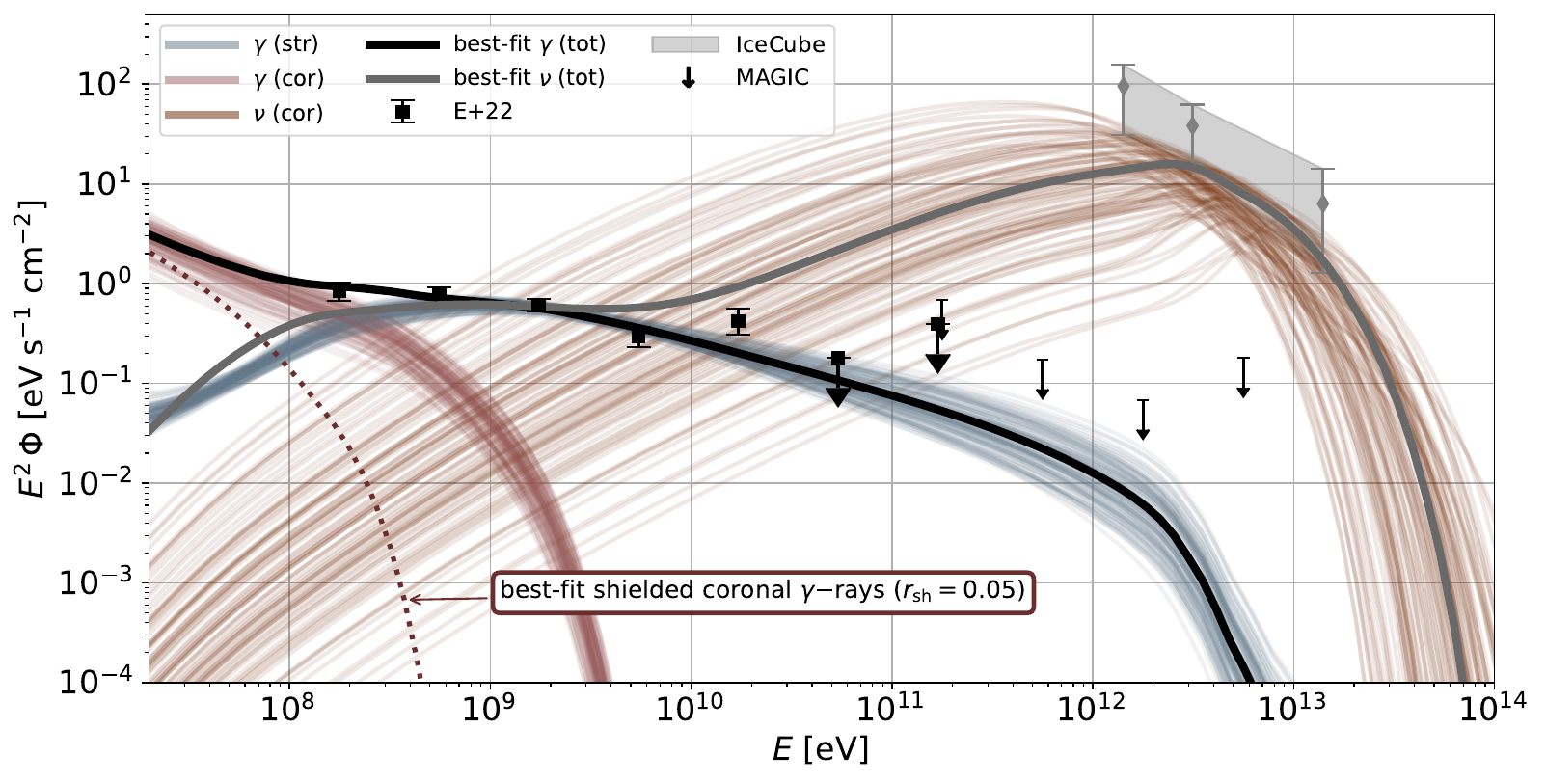}
    \caption{The best-fit model prediction (and 100 alternative scenarios within 1$\sigma$ of the best-fit in transparent colored lines) of the \emph{photon ($\gamma$) and neutrino ($\nu$)} SED of NGC\,1068. The grey markers indicate characteristic neutrino flux values within the observed 68\% confidence band that are used in the MCMC.}
    \label{fig:SED_NGC1068} 
\end{figure*}

As shown in Fig.~\ref{fig:SED_NGC1068}, the updated Seyfert-starburst composite model is still able to explain the non-thermal observational data of NGC\,1068. Such as in E+22, the GeV $\gamma$-rays originate from the starburst-ring, whereas with decreasing energy an increasing percentage of the observed sub-GeV $\gamma$-rays result from the AGN corona. 
%We noticed the first 4FGL\change{-DR4} \citep{4fgl} data point \citep{Abdollahi+2022} 
We noticed the first Fermi-LAT data point\footnote{\change{From private communication with Emma Kun --- more details can be found in Table~\ref{tab:fitData1068}.}}
has a significant impact on the resulting solution, but at these energies the adopted spatial homogeneity has also a strong impact on the $\gamma\gamma$-pair attenuation as indicated by the black dotted line (that illustrates the result for a shield thickness of 5\%). For most scenarios, the predicted neutrino flux is about a factor of two too small (but still partially within the uncertainty band) compared to the IceCube data. The overall spectral behavior between 1.4 and 14 TeV agrees in many cases quite well. Adopting that the high-energy emission is also surrounded by an isotropic X-ray target would enable an increased $\gamma$-ray and neutrino emissivity leading to a better agreement with the IceCube data. Due to the comparably large uncertainties of the IceCube data, the neutrino flux predictions within the $1\sigma$ band of the best-fit show a huge variety, whereas the mm-flux predictions hardly change. Therefore, we also investigate the case with an increased uncertainty of the mm-data --- by using a minimal uncertainty of $\gtrsim 30\,\%$ --- which reduces the need to obtain a coronal synchrotron bump at $100\,\text{GHz}$, so that a bump at slightly higher frequencies is sufficient. This enables higher values of $\omega_{\rm T}$ --- which shifts the characteristic synchrotron self-absorption frequency towards higher frequencies as systematically shown by \cite{delPalacio2025} and linearly increases the target gas density needed for the neutrino production --- and smaller values of $\beta_{\rm B}$ --- which ensures that the necessary CR proton energy for the high energy neutrino production can still be reached --- leading to a perfect agreement with the IceCube data, as shown in Fig.~\ref{fig:SED_NGC1068_err30}.

The most striking parameter result, with respect to the prior expectation, is the spectral index $s$ of the source rate in the starburst region, which needs to be $s=-2.26\pm 0.15$ according to the posterior distributions (see Fig.~\ref{fig:corner_NGC1068}). This outcome supports that either the CR particles escape from their accelerators independent of their energy or acceleration and radiation losses occur in the same environment, i.e.\ the starburst ring. All other parameters are close to the non-uniform prior expectations. For the coronal radius, we obtain a rather distinct range of $\log(r_{\rm c})=1.54\pm 0.06$, which is in perfect agreement with the previous findings from \cite{mutie2025consistent}.
%, except for the coronal turbulence strength $\eta^{-1}_{\rm c}$ that ...
%%%%%%%%%%%%%%%%%%%%%%%%%%%%%%%%%%%%%%%%%%%%%%%%%
%%%%%%%%%%%%%%%%%%%%%%%%%%%%%%%%%%%%%%%%%%%%%%%%%
\subsection{NGC\,7469}
\begin{figure*}
    \centering
    \includegraphics[width=0.48\linewidth]{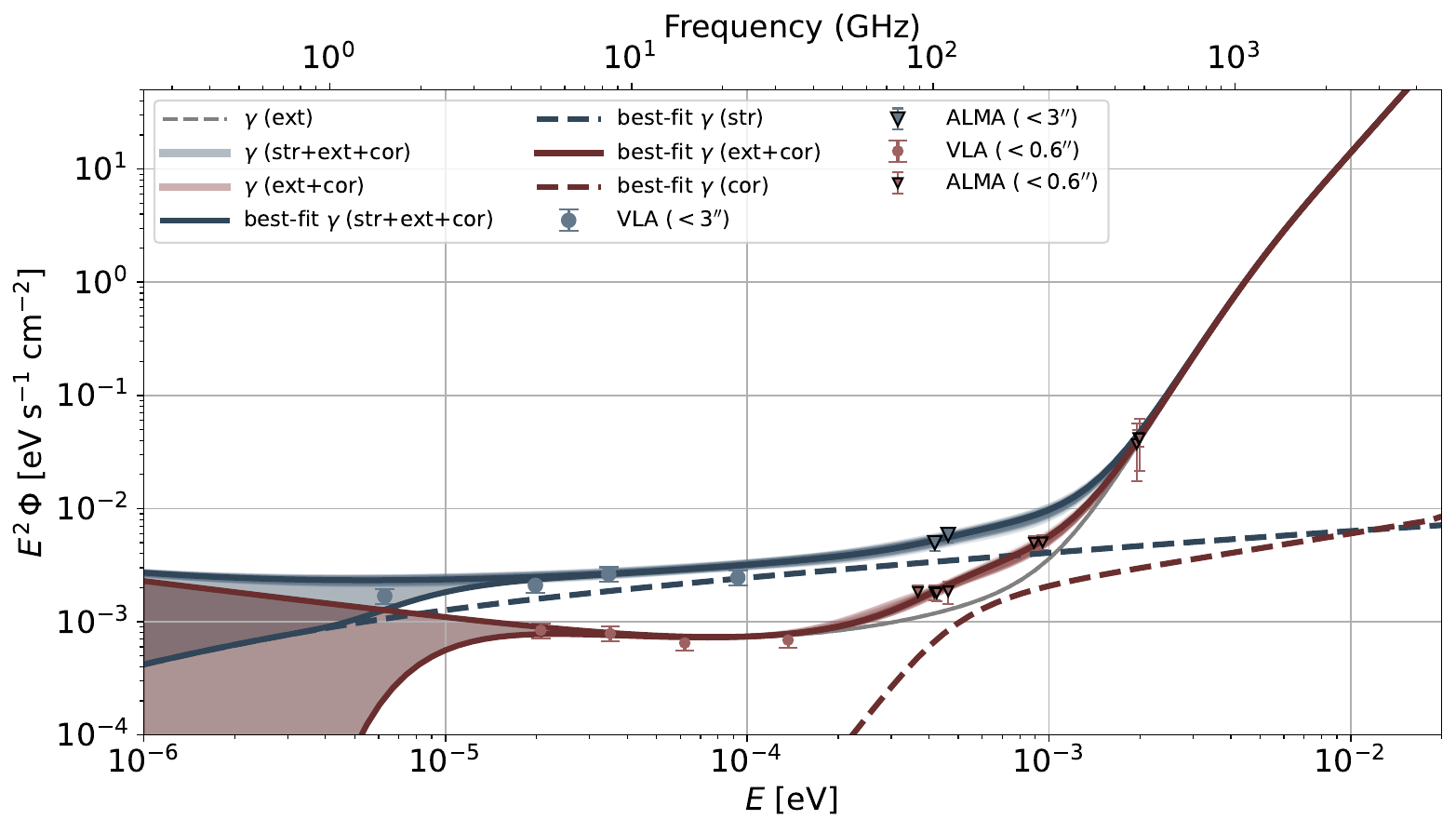}
    \includegraphics[width=0.48\linewidth]{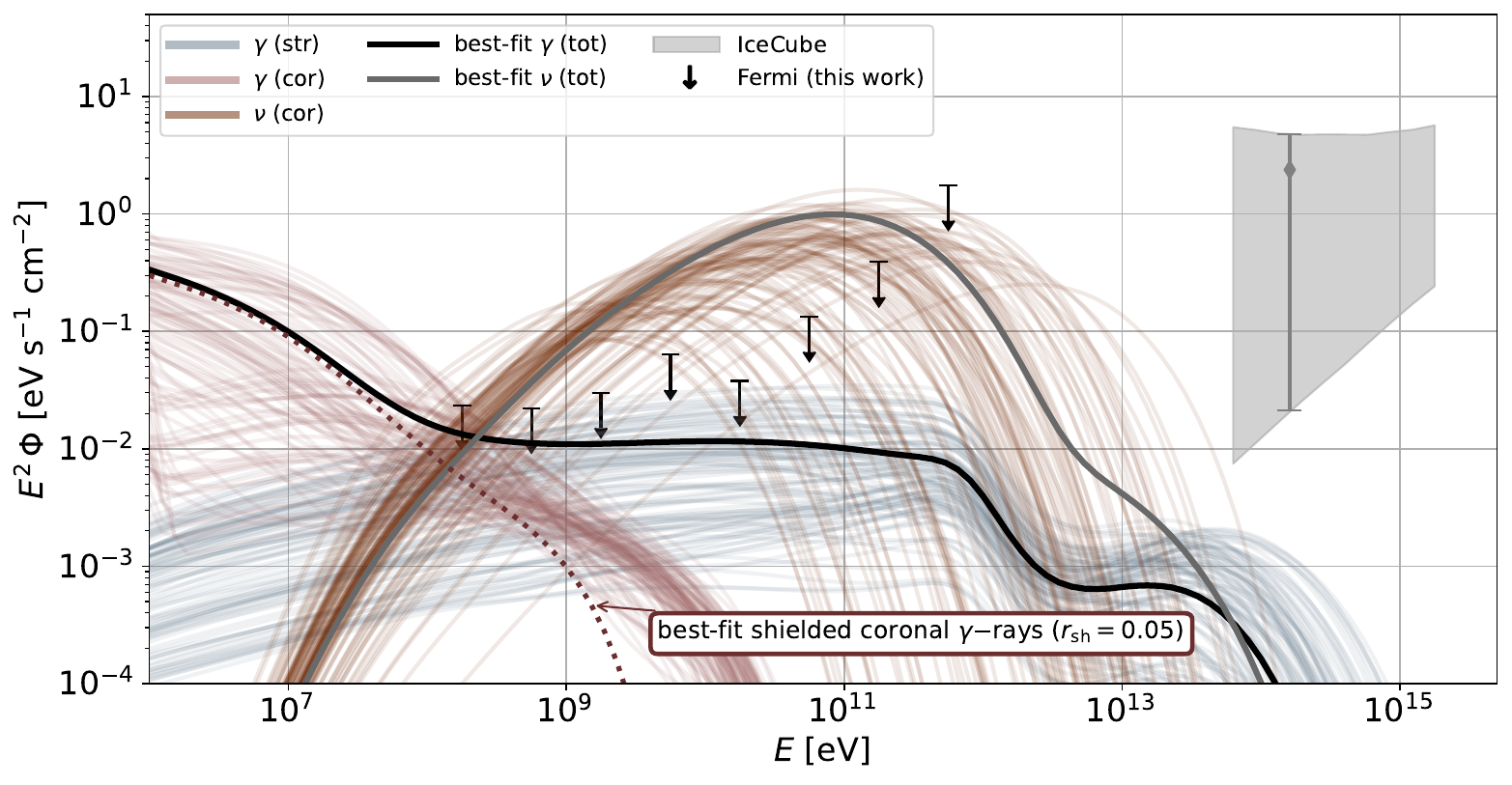}
    \caption{The best-fit model predictions (and 100 alternative scenarios within 1$\sigma$ of the best-fit in transparent colored lines) of the \emph{photon ($\gamma$) and neutrino ($\nu$)} SED of NGC\,7469. The blue and red shaded bands at low energies refer to the uncertainty of the adopted extended gas distribution and its opacity with respect to free-free absorption. The grey marker indicate the characteristic neutrino flux value of the average energy \citep[weighted with the signalness from][]{sommani2025two} of the two
neutrinos within the confidence band that is used in the MCMC.}
    \label{fig:SED_NGC7469} 
\end{figure*}

For the high-energy emission of NGC\,7469 we first verified the basic result from \cite{Salvatore+2025_arXiv}, showing that even a neutrino flux of $(32\pm 16)$~eV\,s$^{-1}$\,cm$^{-2}$ at 161\,TeV --- based on using two associated neutrinos in 4 years of data \citep[see][]{sommani2025two} --- can be explained by the AGN corona without overshooting the $\gamma$-ray upper limits (see Fig.~\ref{fig:NGC7469_SED_onlyHE}). Here, a small coronal radius on the order of a $\text{few}\times \mathcal{R}_{\rm s}$ is needed. However, the mm-data also introduce the need for a coronal contribution in the radio band, which is subsequently taken into account. The resulting best-fit scenario (see Fig.~\ref{fig:SED_NGC7469}) indicates that this mm-bump can be produced by the corona if it extends on scales of $>100\,\mathcal{R}_{\rm s}$. But in that case there is no significant neutrino emission at some tens or hundreds of TeV. So, a spatial homogeneous corona can either explain the mm-bump or the observed hints for TeV-neutrinos, but not both. Note that even in the case of a  minimal uncertainty of $>30\%$ the resulting best-fit scenario does not explain the current hints for high-energy neutrino emission due to the $\gamma$-ray constraints.

With respect to the starburst ring, our model can explain the large scale radio emission and yields a GeV $\gamma$-ray flux that is just about a factor of two below the current upper limits.\footnote{In total the resulting sub-GeV $\gamma$-ray flux is for most scenarios even less than a factor of two.} The associated posterior distributions (see Fig.~\ref{fig:corner_str_NGC7469}) indicate that $s=-1.5\pm 0.2$ is expected, which is significantly closer to the expectation as for NGC\,1068. Further, a slightly higher magnetic field strength as expected of $B_{\rm s} > 10^{-3.7}\,\text{G}$ is needed to explain the large-scale radio data by synchrotron radiation from predominantly primary electrons. This results from the constrain on the maximal IC $\gamma$-ray emission, which yields an upper limit on the non-thermal electron density, and hence, only leaves the magnetic field strength (apart from the spectral index of the CRs) to steer the primary synchrotron flux close to the data. All of the other parameters are in good agreement with the prior expectation. 
%very close to the prior expectation; only the gas density shows a slightly smaller value of $\log(n_{\rm gas}/1\,\text{cm}^{-3})=???\pm ???$ than .

%%%%%%%%%%%%%%%%%%%%%%%%%%%%%%%%%%%%%%%%%%%%%%%%%%%%%%%%%%%%%%%%%%%%%%%%%%%%%%%%%%%%%%%
%%%%%%%%%%%%%%%%%%%%%%%%%%%%%%%%%%%%%%%%%%%%%%%%%%%%%%%%%%%%%%%%%%%%%%%%%%%%%%%%%%%%%%%
%%%%%%%%%%%%%%%%%%%%%%%%%%%%%%%%%%%%%%%%%%%%%%%%%%%%%%%%%%%%%%%%%%%%%%%%%%%%%%%%%%%%%%%
\section{Conclusions and Discussion} \label{sec:conclusions}
In this work, we update the steady-state Seyfert-starburst composite model of E+22 by a more detailed treatment of its AGN corona: 
\begin{enumerate}
    \item[(i)] SDA and energy losses at work, which yields a significant hardening of the resulting non-thermal proton distribution.
    \item[(ii)] Inclusion of the non-linear secondary $e^\pm$ contribution, i.e. from leptonic radiation processes, that however, becomes only relevant below about some hundreds of MeV.
    \item[(iii)] Inclusion of the $\gamma\gamma$-pair attenuation of the escaping $\gamma$-rays as well as a potential shielding scenario (where the X-ray corona is extended beyond the spatial scales of the $\gamma$-ray production site) has a strong impact on the resulting sub-GeV $\gamma$-ray prediction if the radius of the X-ray corona is $\ll 100\mathcal{R}_{\rm s}$. 
\end{enumerate} 
Hereby, we fix the magnetic turbulence and adopt that the non-thermal particles draw a sufficiently small amount of energy from the turbulent flow, so that no significant damping of the cascade occurs. According to \cite{LemoineMuraseRieger2024} damping is inefficient for $(v_{\rm A}/c)^{q-1}\mathcal{E}_{\rm nth}/\mathcal{E}_{\rm B_0}\ll 1$, where $\mathcal{E}_{\rm nth}/\mathcal{E}_{\rm B_0}$ denotes the ratio of the total non-thermal particle energy density over the magnetic energy density. We verified that for the given best-fit scenarios of NGC\,1068 and NGC\,7469 this condition is fulfilled. Note, that if the turbulent cascade is not sufficiently powered externally, the stochastic particle acceleration will become nonlinear and the resulting non-thermal particle energy spectrum softens significantly at high energies, as shown by \cite{LemoineMuraseRieger2024,LemoineRieger2025}. 

We subsequently used an MCMC method to study the parameter space of the Seyfert-starburst composite galaxies NGC\,1068 and NGC\,7469. Such as in E+22, but with a more moderate CR pressure of about $0.1\,P_{\rm gas}$, we are able to explain the multimessenger data of NGC\,1068. However, a minor deficit (of about a factor of two) in the high-energy neutrino flux is obtained, if we explain the observed mm-bump accurately. Since the likelihood is typically dominated by the radio data, we obtain in that case a uniform coronal radius of about $35\,\mathcal{R}_{\rm s}$, which is in perfect agreement with the previous findings from \cite{mutie2025consistent}. With respect to the neutrino deficit it needs to be taken into account that the neutrino flux measurement adopts a power-law
with a best-fit spectral index of $-3.2\pm0.2$. Using a different spectral behavior with a higher peak flux could decrease the observed flux above 1\,TeV, such as shown in \cite{Desai+2025}. 
Moreover, we adopted that the coronal radio emission emerges from the same spatial region such as the high-energy $\gamma$-ray and neutrino emission. But in principle, the confinement of charged particles depends on their Larmor radius, hence, the coronal radius for those particles that emit in the radio band could be much larger than the coronal radius at $\gamma$-ray energies. The MCMC results indicate that such a differentiation, where the radio corona is significantly larger than the $\gamma$-ray corona would improve the fit to the data. 

In the case of NGC\,7469 the MCMC fits indicate clearly the need for a spatially inhomogeneous AGN corona: The mm-bump needs to emerge from a coronal size $R_{\rm c}>100\,\mathcal{R}_{\rm s}$, whereas neutrino emission at energies $>100\,\text{TeV}$ requires $R_{\rm c}< 10\,\mathcal{R}_{\rm s}$ if there is no shielding outer photon target. A homogeneous corona with a fixed size at all energies can explain the features either at high energies or in the radio band, but not both. This discrepancy is generally known from other wavebands, e.g.\ the sizes of the coronae inferred from X-rays is about $R_{\rm c}\lesssim 10\,\mathcal{R}_{\rm s}$ \citep{Fabian2015}, whereas the size of the mm emitter inferred via modeling of the spectral energy distribution (SED) yields sizes that are often larger, $R_c \sim (60$--$250)\,\mathcal{R}_{\rm s}$ \citep{delPalacio2025}. 
We also note that the size discrepancy is too large to be explained just by time variability, with even the most extreme case of mm variability reported in \cite{Shablovinskaya2024} corresponding to only a change in size of about a factor of two. 
% We note that temporal variability could also add to this size discrepancy, as mm variability has been reported in some AGN \citep[e.g.][]{Shablobinskaya2024}, and NGC\,7469 exhibits X-ray variability \citep{mehdipour2018multi}. However, the size discrepancy is likely too large to be explained just by variability, with the most extreme case reported in \citep{Shablovinskaya2024} the change in size was about a factor of two.

Moreover, the predicted best-fit GeV-$\gamma$-ray flux of the starburst ring is just about a factor of two smaller than the current upper limits. In addition, a potentially dominant $\gamma$-ray contribution could emerge at sub-GeV energies, where future observation will be able to draw further constraints.
%%%%%%%%%%%%%%%%%%%%%%%%%%%%%%%%%%%%%%%%%%%%%%%%%%%%%%%%%%%%%%%%%%%%%%%%%%%%%%%%%%%%%%%
%%%%%%%%%%%%%%%%%%%%%%%%%%%%%%%%%%%%%%%%%%%%%%%%%%%%%%%%%%%%%%%%%%%%%%%%%%%%%%%%%%%%%%%
\section*{Acknowledgements}
BE and SS thank J. Dörner for fruitful discussion on the MCMC analysis as well as Emma Kun for the provided Fermi-LAT data analysis of NGC\,1068. We acknowledge funding from the German Science Foundation DFG, within the Collaborative Research Center SFB1491 ``Cosmic Interacting Matters - From Source to Signal". 
S.d.P. acknowledges support from ERC Advanced Grant 789410. We further thank Anna Franckowiak, Giacomo Principe, Filippo D'Ammando and Haocheng Zhang for constructive comments on the paper draft.

The \textit{Fermi} LAT Collaboration acknowledges generous ongoing support from a number of agencies and institutes that have supported both the development and the operation of the LAT as well as scientific data analysis. These include the National Aeronautics and Space Administration and the Department of Energy in the United States, the Commissariat à l’Energie Atomique and the Centre National de la Recherche Scientifique / Institut National de Physique Nucléaire et de Physique des Particules in France, the Agenzia Spaziale Italiana and the Istituto Nazionale di Fisica Nucleare in Italy, the Ministry of Education, Culture, Sports, Science and Technology (MEXT), High Energy Accelerator Research Organization (KEK) and Japan Aerospace Exploration Agency (JAXA) in Japan, and the K. A.Wallenberg Foundation, the Swedish Research Council and the Swedish National Space Board in Sweden. Additional support for science analysis during the operations phase is gratefully acknowledged from the Istituto Nazionale di Astrofisica in Italy and the Centre National d’Etudes Spatiales in France. This work is performed in part under DOE Contract DE-AC02-76SF00515.

\vspace{5mm}

%\software{Some of the results in this paper have been derived using the software packages Numpy \cite{vanDerWalt2011}, Scipy \cite{2020SciPy-NMeth}, Matplotlib \cite{Hunter:2007}, Seaborn \cite{michael_waskom_2017_883859}. }

This paper makes use of the following ALMA data: 
ADS/JAO.ALMA\#2017.1.00078.S, % B3 at ~0.2" and 1.5" resolution; also B6 at ~0.3" resolution; also B7 at ~0.3" resolution; also B8 at ~0.3" resolution
ADS/JAO.ALMA\#2013.1.00218.S, % B3 at ~0.5" resolution
ADS/JAO.ALMA\#2013.1.00700.S. %B3 at ~1.5" resolution
ALMA is a partnership of ESO (representing its member states), NSF (USA) and NINS (Japan), together with NRC (Canada), NSTC and ASIAA (Taiwan), and KASI (Republic of Korea), in cooperation with the Republic of Chile. The Joint ALMA Observatory is operated by ESO, AUI/NRAO and NAOJ. %In addition, publications from NA authors must include the standard NRAO acknowledgement: The National Radio Astronomy Observatory is a facility of the National Science Foundation operated under cooperative agreement by Associated Universities, Inc.
%%%%%%%%%%%%%%%%%%%%%%%%%%%%%%%%%%
%%%%%%%%%%%%%%%%%%%
\bibliographystyle{aa} % style aa.bst
\bibliography{references} % your references Yourfile.bib
%%%%%%%%%%%%%%%%%%%%%%%%%%%%%%%%%
\begin{appendix}

\section{Posterior Distributions}\label{sec:posterior_dist}
\begin{figure*}
    \centering
    \includegraphics[width=0.98\linewidth]{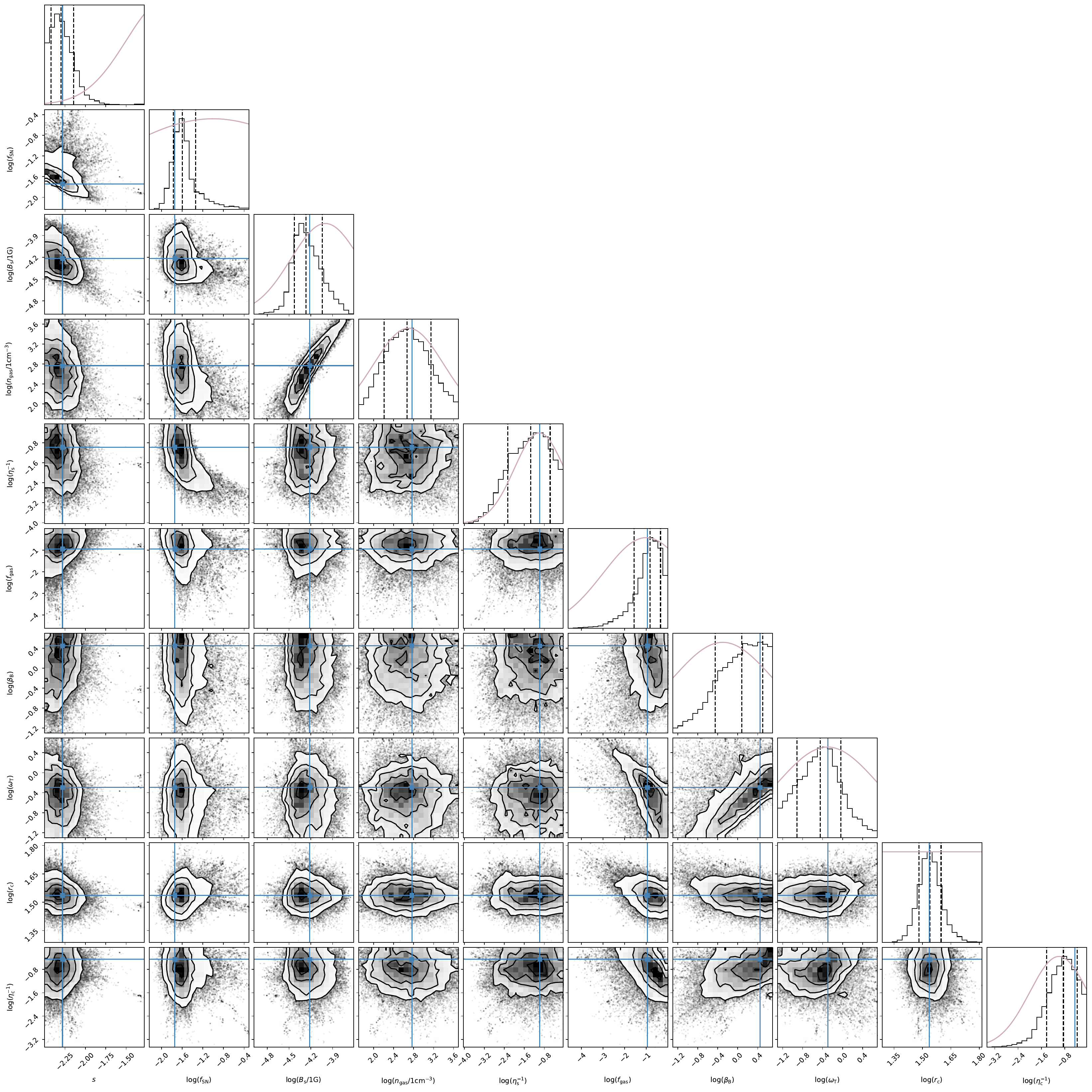}
    \caption{The posterior distribution of the fit parameters for NGC\,1068. The blue lines indicate the best-fit parameter values and the magenta curve illustrates the adopted prior distribution.}
    \label{fig:corner_NGC1068} 
\end{figure*}

The Fig.~\ref{fig:corner_NGC1068} show the posterior distribution of the analyzed parameter space of the starburst ring (first five parameter on the x-axis) and the AGN corona (second five parameters on the x-axis) of NGC\,1068. The associated MCMC fit includes both environments, as they cannot be disentangled at high energies. The plot displays the results from $11279\times 30$ iterations, whereof the first 800 of each of the 30 walkers are due to correlation effects excluded, and only every 5th outcome is used.

\begin{figure*}
    \centering
    \includegraphics[width=0.95\linewidth]{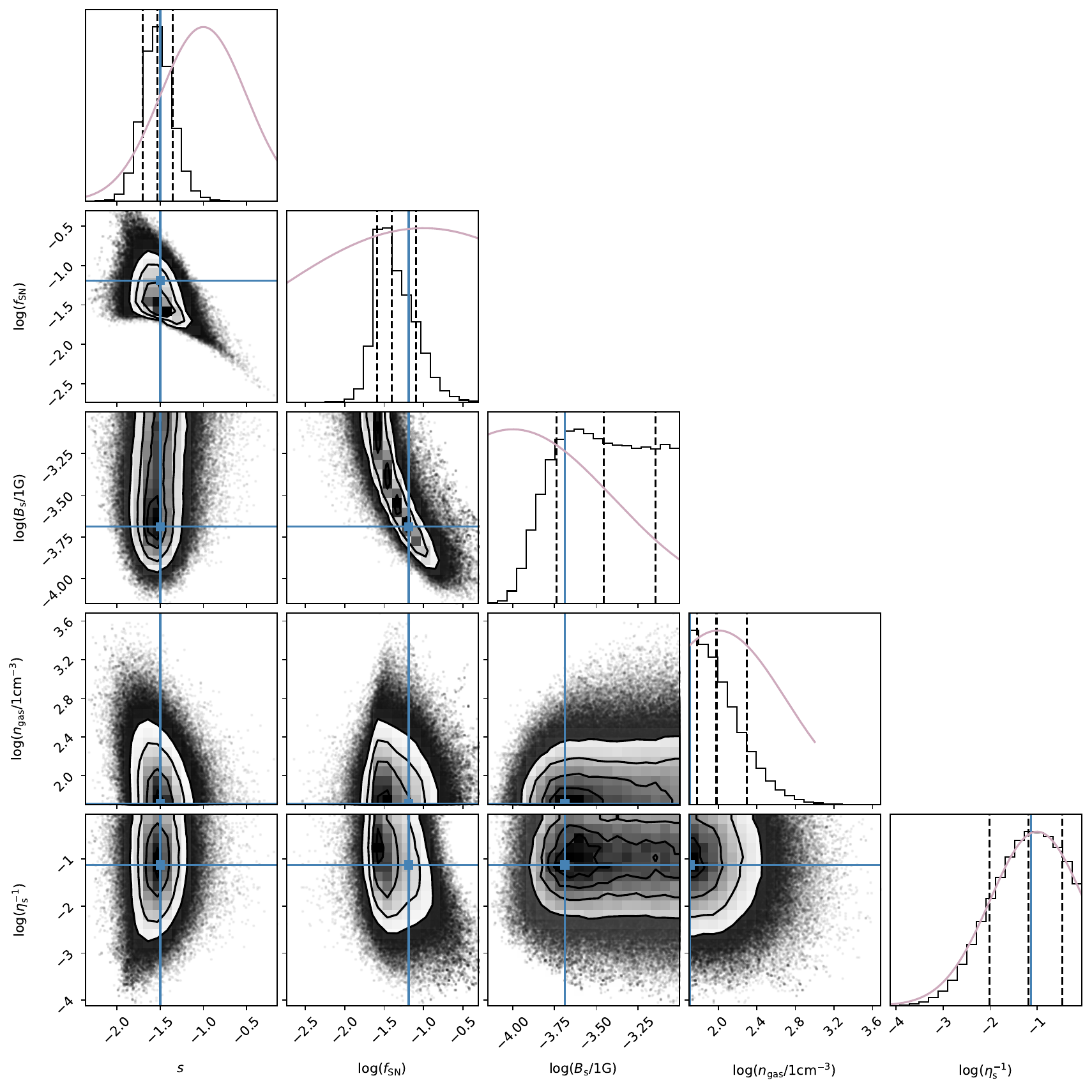}
    \caption{The posterior distribution of the fit parameters for the starburst ring of NGC\,7469. The blue lines indicate the best-fit parameter values and the magenta curve illustrates the adopted prior distribution. }
    \label{fig:corner_str_NGC7469} 
\end{figure*}

In contrast to the previous case the MCMC fit of NGC\,7469 has been performed for the AGN corona and the starburst ring separately, as there are currently only upper limits in the $\gamma$-ray band available. Therefore, the main contributions from those two environments can be nicely distinguished, if the high-energy neutrino hints emerge predominantly from the corona. Since the AGN corona model cannot explain the associated date, we only present the parameter analysis of the starburst ring (see Fig.~\ref{fig:corner_str_NGC7469}). Here, $15000\times100$ iterations have been performed, whereof only every fifth of the final $14000\times100$ iterations is used.

In all MCMC runs we have verified that the calculations have converged.

\section{Additional Results}\label{sec:add_outcome}
\begin{figure*}
    \centering
        \centering
    \includegraphics[width=0.48\linewidth]{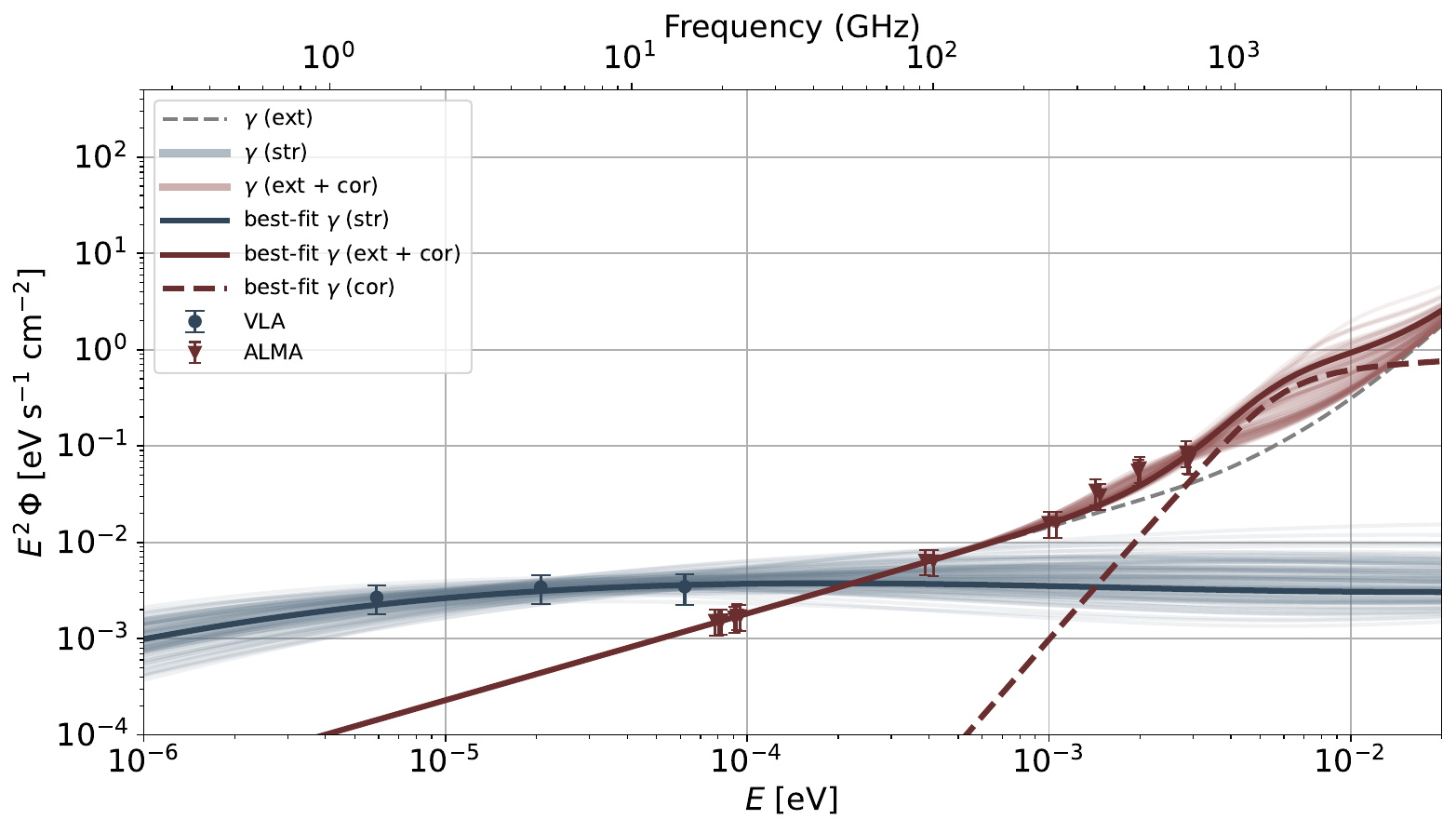}
    \includegraphics[width=0.48\linewidth]{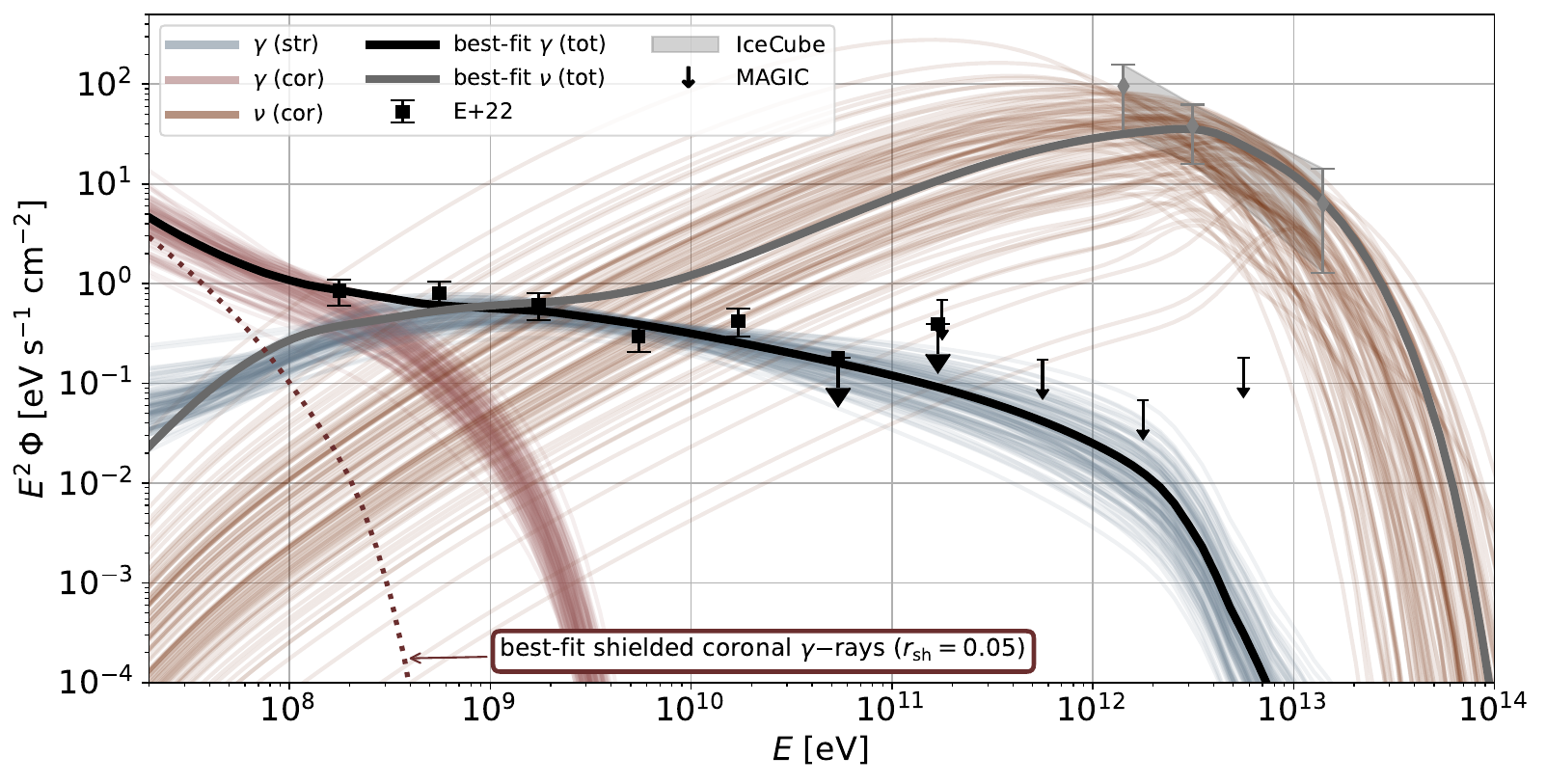}
    \caption{The best-fit model prediction (and 100 alternative scenarios within 1$\sigma$ of the best-fit in transparent colored lines of the \emph{photon ($\gamma$) and neutrino ($\nu$)} SED of NGC\,1068 assuming a minimal data uncertainty of 30\,\%. The grey markers indicate characteristic neutrino flux values within the observed 68\% confidence band that are used in the MCMC.}
    \label{fig:SED_NGC1068_err30} 
\end{figure*}

For comparison, we relax the observational constrain by the mm data in the likelihood by introducing a minimal relative uncertainty of 30\%. This yields a perfect agreement with the high-energy neutrino data, as shown in Fig.~\ref{fig:SED_NGC1068_err30}. The most striking difference in the associated fit parameters with respect to the ones from Fig.~\ref{fig:corner_NGC1068}, is a reduced coronal radius of about $28\,\mathcal{R}_{\rm s}$ and an increased optical Thomson depth of $2$. Hence, the production of high-energy neutrinos in a more compact, denser region within the X-ray corona could lead to a significant improvement of the fit.

\begin{figure*}
    \centering
        \centering
    \includegraphics[width=0.49\linewidth]{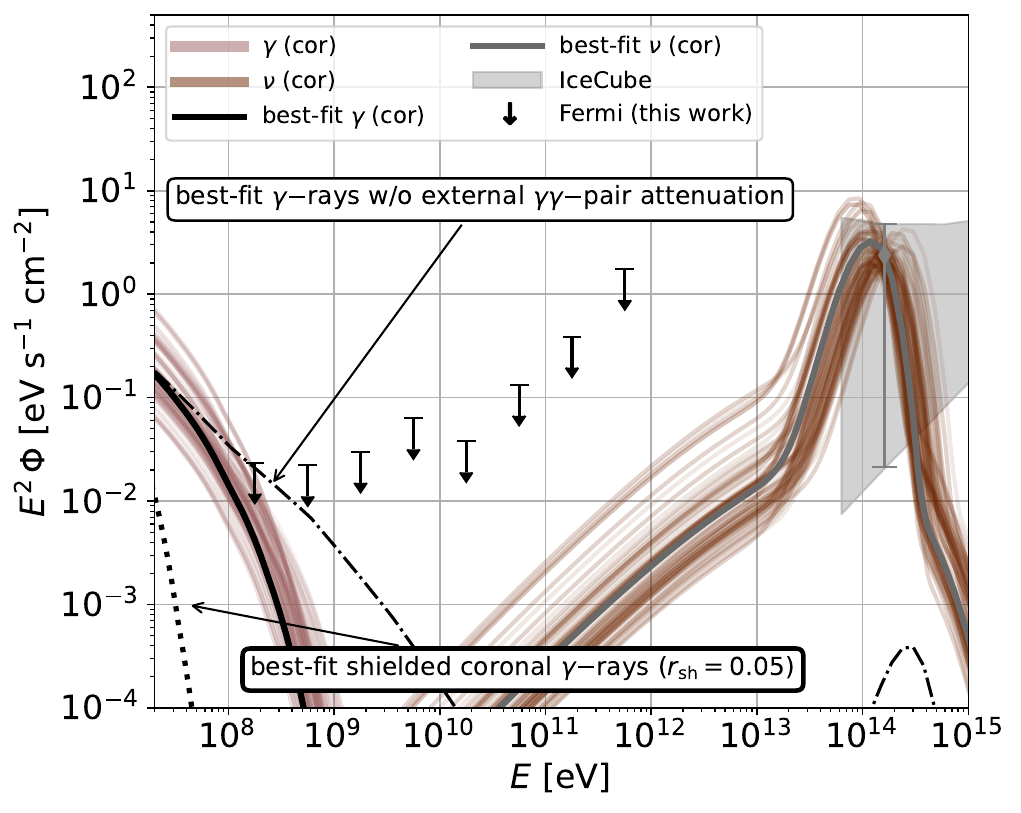}%{figures/NGC7469_SED_onlyHE.pdf}
    \includegraphics[width=0.49\linewidth]{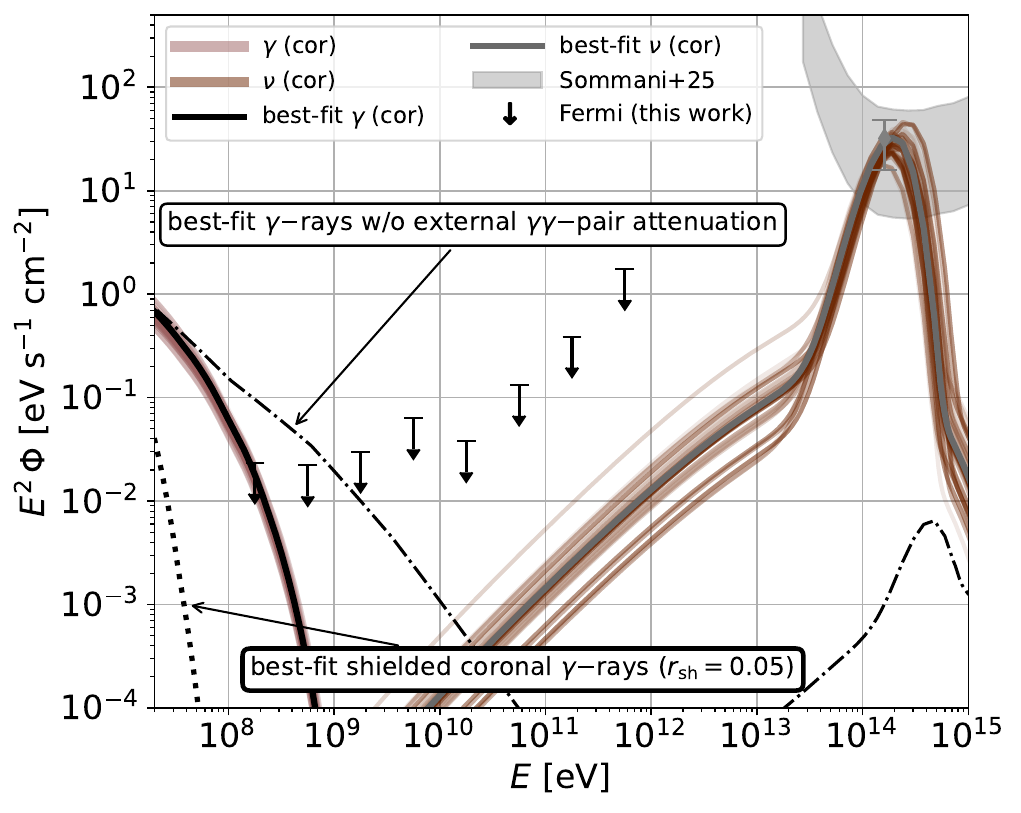}
    \caption{The best-fit model predictions (and 100 alternative predictions with only a marginally different posterior probability in transparent colored lines) of the coronal \emph{photon ($\gamma$) and neutrino ($\nu$)} SED of NGC\,7469 at high energies considering only the $\gamma$-ray upper limits and the neutrino expectation from \cite{IceCube2025_arxiv} (\emph{left}) and \cite{sommani2025two} (\emph{right}), respectively. For the latter case (right), the 90\% confidence interval (grey band) is converted to a 68\% confidence value (grey marker) at the weighted average energy of 161 TeV, which yields $(32\pm 16)\,\text{eV}\,\text{cm}^{-2}\,\text{s}^{-1}$ for 4 years of data and two associated muon neutrinos.}
    \label{fig:NGC7469_SED_onlyHE} 
\end{figure*}

For NGC\,7469 we also tested to fit only the high-energy emission by the AGN corona. Here, we used only the upper limits on the gamma-ray flux as well as the neutrino flux indications at 161\,TeV using the IceCube analysis given in \cite{IceCube2025_arxiv} (\emph{left} plot in Fig.~\ref{fig:NGC7469_SED_onlyHE}) and the two-neutrino estimate from \cite{sommani2025two} (\emph{right} plot in Fig.~\ref{fig:NGC7469_SED_onlyHE}), respectively. It is shown that that a perfect agreement with the hints for neutrino emission is possible, even in the case of the higher flux expectation from \cite{sommani2025two}. But for this purpose the high-energy emission needs to emerge from an extremely compact coronal region of $\ll 10\,\mathcal{R}_{\rm s}$. Note that due to the lack of observational constraints at high energies, the MCMC is strongly guided by the prior information leading to a huge number of predictions within the $1\sigma$-regime of the best-fit scenario that, however, do not properly explain the observational hints for neutrino emission. Therefore, we decided to display only a few of those predictions with a posterior probability that deviates only marginally (i.e.\ $\ll 1\sigma$) from the best-fit.
%Note that in principle also few $\times\,\mathcal{R}_{\rm s}$ is possible to explain the neutrino flux with no significant $\gamma$-ray flux above about 100\,MeV.
\end{appendix}
%%%%%%%%%%%%%%%%%%%%%%%%%%%%%%%%%%%%%%%%%%%%%%%%%%%%%%%%%%%%%%%%%%%%
\end{document}